\documentclass[twocolumn,twocolappendix]{aastex631}

\usepackage{amsmath, amssymb}
\usepackage{graphicx} \usepackage{footnote}
\usepackage{url}

\usepackage{soul} \usepackage{xcolor}

\usepackage{hyperref}
\usepackage{times}
\usepackage{beramono}

\defcitealias{Dejonghe1988}{DZ88}
\defcitealias{Sellwood1997}{SV97}

\newcommand\soutpars[1]{\let\helpcmd\sout\parhelp#1\par\relax\relax}

\usepackage{acronym}

\newcommand{\p}{\partial}
\newcommand{\sgn}{\mathrm{sgn}}

\newcommand{\rd}{\mathrm{d}}
\newcommand{\re}{\mathrm{e}}
\newcommand{\ri}{\mathrm{i}}
\newcommand{\rs}{\mathrm{s}}

\newcommand{\rr}{\mathrm{r}}
\newcommand{\rT}{\mathrm{T}}
\newcommand{\rD}{\mathrm{D}}
\newcommand{\stak}{St{\"a}ckel}
\newcommand{\eff}{\mathrm{eff}}
\newcommand{\deltaD}{\delta_{\rD}}

\newcommand{\Ftot}{F_{\mathrm{tot}}}

\newcommand{\br}{\boldsymbol{r}}

\newcommand{\bw}{\boldsymbol{w}}

\newcommand{\btheta}{\boldsymbol{\theta}}

\newcommand{\brc}{\mathbf{c}}

\newcommand{\brg}{\mathbf{g}}
\newcommand{\brA}{\mathbf{A}}

\newcommand{\bJ}{\boldsymbol{J}}

\newcommand{\Lz}{L_z}
\newcommand{\Jr}{J_{\rr}}

\newcommand{\alphar}{\alpha_{\rr}}

\newcommand{\rM}{\mathrm{M}}
\newcommand{\bM}{\mathbf{M}}
\newcommand{\bI}{\mathbf{I}}
\newcommand{\bE}{\mathbf{E}}

\newcommand{\bk}{\boldsymbol{k}}
\newcommand{\tbk}{\widetilde{\bk}}
\newcommand{\bOmega}{\boldsymbol{\Omega}}

\newcommand{\tF}{\widetilde{F}}

\newcommand{\tE}{\widetilde{E}}
\newcommand{\tLz}{\widetilde{L}_{z}}
\newcommand{\ta}{\widetilde{a}}
\newcommand{\tc}{\widetilde{c}}

\newcommand{\Frot}{F_{\mathrm{rot}}}

\newcommand{\bfct}{\boldsymbol{j}}

\usepackage{soul} \definecolor{aquamarine}{rgb}{0.5, 1.0, 0.83}
\definecolor{softblue}{rgb}{0.5, 0.8, 0.93}

\begin{document}

 \title{  Linear response of  rotating and flattened  stellar clusters:
the oblate Kuzmin--Kutuzov \stak\ family.
 }

 \author[0009-0002-8012-4048]{Kerwann Tep}
\affiliation{Department of Physics and Astronomy,
    University of North Carolina at Chapel Hill,
    120 E. Cameron Ave, Chapel Hill, NC, 27599, USA
}\affiliation{Institut d'Astrophysique de Paris,
    CNRS and Sorbonne Université, UMR 7095, 98 bis Boulevard Arago, F-75014 Paris, France
}

\author[0000-0003-0695-6735]{Christophe~Pichon}
 \email{Corresponding author: pichon@iap.fr}
\affiliation{Institut d'Astrophysique de Paris,
    CNRS and Sorbonne Université, UMR 7095, 98 bis Boulevard Arago, F-75014 Paris, France
}\affiliation{IPhT, DRF-INP, UMR 3680, CEA, L'Orme des Merisiers, B\^at 774, 91191 Gif-sur-Yvette, France
}\affiliation{Kyung Hee University, Dept. of Astronomy \& Space Science, Yongin-shi, Gyeonggi-do 17104, Republic of Korea
}

 \author[0000-0003-1517-3935]{Michael S. Petersen}
\affiliation{Institute for Astronomy, University of Edinburgh, Royal Observatory, Blackford Hill, Edinburgh EH9 3HJ, UK
}

\begin{abstract}

This paper investigates the linear response of a series of spheroidal stellar clusters, 
the Kuzmin--Kutuzov \stak\ family, 
which exhibit a continuous range of flattening and rotation, extending from an isochrone sphere to a Toomre disk. The method successfully replicates the growing modes previously identified in published $N$-body simulations. It  relies on the efficiency of the matrix method to quantify systematically the effects of rotation  and flattening
on the eigenmodes of the galaxy. We identify two types of bi-symmetric instabilities for the flatter models -- the so-called  bending and bar-growing modes -- the latter of which persists even for very round models. As anticipated, in its least unstable configurations, the system becomes flatter as its rotational speed increases.
More realistic equilibria will be required to achieve a better match to  
the main sequence of fast-slow rotators.
The corresponding code is made public.

\end{abstract}


\section{Introduction}
\label{sec:intro}

Rotation is ubiquitous in the universe \citep{10.1093/mnras/278.1.27,10.1111/j.1365-2966.2005.09981.x,Bianchini2018}. The diversity among galaxies showcases a broad spectrum of mass and angular momentum distributions \citep{2012ApJS..203...17R,2015ApJ...812...29T}. During formation, gravitational forces drive these celestial bodies to collapse, acquiring angular momentum via torques  \citep{1987ApJ...319..575B}.  Thanks to  dissipation,  the result post virialization  is typically a flattened rotating structure \citep{1980MNRAS.193..189F}. Understanding how  this geometry and kinematics affects the  response of galaxies is important in its own right \citep[e.g., to define stability thresholds, starting with the seminal work of][]{1973ApJ...186..467O}, but also to explain their long-term   evolution through adiabatic  or resonant relaxation \citep{1964AnAp...27...83H,1988MNRAS.230..597B}.
Such endeavor has been attempted through $N$-body simulations \citep[e.g.][]{Palmer1990,Kuijken1994,Sellwood1997,Breen2021,10.1093/mnras/stac2281}\footnote{ We shall refer to \citet{Sellwood1997} as \citetalias{Sellwood1997}.}. 
A worthy alternative is to follow the path of \cite{1977ApJ...212..637K,Polyachenko1981,Saha1991,Weinberg1994} and compute the linear response of such systems \citep[see, e.g.][ for a recent public distribution of such codes for thin discs and spheres]{Petersen2024}.  The knowledge of such response is useful per say \citep{Rozier2022}, but also to quantify their long term evolution:
their secular response will be amplified  by the square of their corresponding gravitational susceptibility, which can be large when centrifugal support and flattening is important  \citep{2001MNRAS.328..321W,2015A&A...584A.129F}. This is the case for the majority of galaxies 
across cosmic time \citep{Nair_2010}.

The impact of  rotation and flattening on their linear response   is most effectively studied analytically  by considering integrable equilibria as a reference point, because it yields trivial unperturbed equations of motions
 \citep{1976ApJ...205..751K,Zeeuw1986,Robijn1996}. Historically, the complexity of moving beyond simple spherical symmetry posed significant challenges (six-dimensional  phase space fully coupled via self-gravity), which were first addressed  in \cite{1995PhDT.......119R} in the so-called thin shell approximation. 
 However, modern computers can now model   more complex shapes or kinematics \citep{Rozier2019}, opening the prospects of also extending our understanding  beyond the  spherical or razor thin geometries.
This motivates the present investigation.

This paper  studies the linear response of a family of spheroids with varying levels of flattening and rotation,  relying on the integrability of oblate \stak\, clusters  \citep[][hereafter \citetalias{Dejonghe1988}]{Zeeuw1986,Dejonghe1988}.
It extends \cite{1995PhDT.......119R} in that it is not limited to shell orbits and allows for rotation.

Section~\ref{sec:prolate_spheroidal} presents the spheroidal coordinate system and derives angles actions for the Kuzmin--Kutuzov \stak\, family of clusters.
Section~\ref{sec:linear_theory} presents the corresponding linear response theory.
We  then
compute the growing modes of sequences   
with varying levels of flattening (Section~\ref{sec:KK_response})   and rotation (Section~\ref{sec:rotation}).  
Section~\ref{sec:discussion} wraps up and discusses prospects.

\section{The oblate spheroidal cluster}
\label{sec:prolate_spheroidal}

Let us review some properties of axisymmetric \stak\ potentials, following  \citet{Zeeuw1986} and \citetalias{Dejonghe1988} (section~II.a).

\subsection{Prolate spheroidal coordinates}
\label{subsec:prolate_coordinates}

We define spheroidal coordinates as the triple $(\lambda,\phi,\nu)$, where $\phi$ is the azimuthal angle from the usual cylindrical coordinates $(R=\sqrt{x^2+y^2},\phi,z)$, and where $\lambda,\nu$ are the roots for $\tau$ of 
\begin{equation}
\label{eq:def_spheroid}
\frac{R^2}{\tau-a^2} + \frac{z^2}{\tau-c^2}=1.
\end{equation}
We shall restrict ourselves to the case
 $a > c$, meaning that the spheroids of constant $\lambda$ are prolate, while the hyperboloids of constant $\nu$ have two
sheets. $\lambda$ and $\nu$ are elliptic coordinates in each meridional plane $\phi=\phi_0$, with foci on the $z$-axis at $z = \pm \Delta$, which 
we define as $\Delta=\sqrt{a^2-c^2}$. We call the coordinates $(\lambda,\nu,\phi)$ the \textit{confocal elliptic coordinates}, from which we can define other useful parametrizations such as the \textit{elliptic coordinates} $(u,v,\phi)$ and $(\xi,\eta,\phi)$. We refer to Appendix~\ref{app:convention_spheroidal} and Fig.~\ref{fig:LambdaNu} for more details.

\subsection{Angle-action coordinates}

The Hamiltonian of the system is given by \citep[see, e.g., equation~3.246 of][]{Binney2008}
\begin{align}
H &\!=\! \frac{p_u^2 + p_v^2}{2\Delta^2 (\sinh^2 u + \sin^2 v)} \!+\! \frac{p_{\phi}^2}{2\Delta^2 \sinh^2 u \sin^2 v} + \psi,\!
\end{align}
where $\psi$ is the potential of the system given by  \cite{Jacobi1866,Zeeuw1985b}:
\begin{align}
\label{eq:psi_stackel}
\psi=\frac{f(\lambda) - f(\nu)}{\lambda-\nu}= \frac{U(u)-V(v)}{\sinh^2 u + \sin^2 v},
\end{align}
and we have the relations
\label{eq:U_V_G}
\begin{align}
U(u)=  \frac{ f(\lambda)}{\Delta^2};\quad
V(v)= \frac{ f(\nu)}{\Delta^2}.
\end{align}
The momenta are then given by
\begin{subequations}
\begin{align}
p_u^2 &= 2 \Delta^2 ( E \sinh^2 u - I_3 - U[u]) - \frac{\Lz^2}{\sinh^2 u},\\
p_v^2 &= 2 \Delta^2 ( E \sin^2 v + I_3 + V[v]) - \frac{\Lz^2}{\sin^2 v},
\end{align}
\end{subequations}
where  $I_3$ is a third integral of motion such that $2 \Delta^2 I_3 \!\rightarrow\! L^2$ as $\Delta \!\rightarrow\! 0$ \citep[see, e.g., equation~3.347 of][for an expression]{Binney2008}. 
In this formulation, we can express the action coordinates \citep[see, e.g., equation~3.250 of][]{Binney2008}\footnote{We show in appendix~\ref{app:Ju_to_Jlambda} that the action variables $(J_u,J_v,\Lz)$ used in the $(u,v)$ formulation are identical to the action variables $(J_{\lambda},J_{\nu},\Lz)$ used in the $(\lambda,\nu)$ formulation.}
\begin{subequations}
\label{eq:defaction}
\begin{align}
J_u &= \frac{1}{\pi} \int_{u_0}^{u_1} \rd u \,p_u,\\
J_v &= \frac{1}{\pi} \int_{v_0}^{v_1} \rd v \,p_v=\frac{2}{\pi} \int_{\pi/2}^{v_1} \rd v \,p_v ,\\
J_{\phi}&= \Lz. 
\end{align} 
\end{subequations}
We refer to appendix~\ref{app:boundaries} for the exact computation of the boundaries of the ``radial" motion, $(u_0, u_1$), and of the out-of-plane motion, $(v_0, v_1)$. Following equation~(13) of \citet{Binney2012}, the associated angle variables are given by
\begin{subequations}
\label{eq:defangles}
\begin{align}
\theta_u &= \int_{u_0}^u \rd u' \frac{\p p_u}{\p J_u} +  \int_{v_0}^v \rd v' \frac{\p p_v}{\p J_u},\\
\theta_v &= \int_{u_0}^u \rd u' \frac{\p p_u}{\p J_v} +  \int_{v_0}^v \rd v' \frac{\p p_v}{\p J_v},\\
\theta_{\phi} &= \int_{u_0}^u \rd u' \frac{\p p_u}{\p \Lz} +  \int_{v_0}^v \rd v' \frac{\p p_v}{\p \Lz} + \phi.
\end{align}
\end{subequations}
One can show that, in the spherical limit, the spheroidal coordinates reduce to the spherical coordinates $(r,\vartheta,\phi)$, such that $\lambda \rightarrow r^2 + a^2$ and $v \rightarrow \vartheta$. Furthermore, $J_u$ tends to the radial action $\Jr$ while $J_v$ tends to the longitudinal action $J_{\vartheta}=L-|\Lz|$ \citep[see, e.g.,][]{Zeeuw1990}.

\section{Linear response theory}
\label{sec:linear_theory}

\subsection{The matrix method}
Let us consider a self-gravitating stellar cluster, and place ourselves in an inertial frame. We assume that we have access to a complete set of bi-orthogonal basis functions $(\psi^{(p)}[\br],\rho^{(p)}[\br])$ which satisfy
\begin{subequations}
\begin{align}
&\psi^{(p)}(\br) = \int \rd \br' U(|\br-\br'|) \rho^{(p)}(\br'),\\
&\int \rd \br \psi^{(p)*}(\br) \rho^{(q)}(\br)=-\delta_{pq},
\end{align}
\end{subequations}
where we defined the Newtonian interaction potential, ${U(r)\!=\!-G/r}$. 

Suppose that the equilibrium is described by a set of angle actions coordinates $(\btheta,\bJ)$ 
given by equations~\eqref{eq:defaction}--\eqref{eq:defangles}, and a distribution function, $F(\bJ)$
given by equation~\eqref{eq:Ftot_KK}  below. Then, its response matrix, $\bM(\omega)$, is defined through its components\footnote{If the frame is non-inertial, then we must take into account inertial pseudo-forces in the computation \citep[see, e.g.,][]{Rozier2022}.} \begin{equation}
\label{eq:Mpq_generic}
\rM_{pq}(\omega)\!=\!(2\pi)^d \sum_{\bk} \!\!\int \!\!\rd \bJ \frac{\bk \cdot \p F/\p \bJ}{\omega \!-\! \bk \cdot \bOmega(\bJ)} \psi_{\bk}^{(p)*}  \psi_{\bk}^{(q)} , 
\end{equation}
where $\mathrm{Im}(\omega)\!>\!0$,
 $\bOmega(\bJ)=\p H_0/\p \bJ$ are the orbital frequencies, $d$ the dimension of physical space and $ \psi_{\bk}^{(p)}(\bJ)$ is the Fourier transform of the potential basis elements defined by
\begin{equation}
 \psi_{\bk}^{(p)}(\bJ) = \int \frac{\rd \btheta}{(2\pi)^d}  \psi^{(p)}(\br[\btheta,\bJ]) \re^{-\ri \bk \cdot \btheta}.
\end{equation}
We detail this computation in Appendix~\ref{app:basis}.

\subsection{Prolate spheroidal basis functions}

Let use write the basis elements in a separable way
\begin{align}
&\psi^{\ell m n}(\br) =\frac{\sqrt{4\pi G }}{\Delta} F^{\ell m n}(\xi) Y_{\ell}^m(v,0)\,\re^{\ri m \phi}.
\end{align}
Let $\bk=(k_{u},k_{v},k_{\phi})$. Then
\begin{align}
\label{eq:FT_basis}
\psi_{\bk}^{\ell m n}(\bJ) 
&=\frac{\sqrt{4\pi G }}{\Delta}\delta_{k_{\phi}}^{m} \frac{1+(-1)^{\ell+m+k_v}}{2} W_{\bk}^{\ell m n}(\bJ),
\end{align}
where \begin{align}
W_{\bk}^{(p)} (\bJ)
&=\int_{u_0}^{u_1} \hspace{-1mm} \frac{ \rd u }{\pi} F^{\ell m n}(\xi)  \cos( \alpha_{\bk} ) \\
&\times\int_{v_0}^{v_1} \hspace{-1mm}  \frac{ \rd v }{ \pi  }  |J |\, Y_{\ell}^m(v,0) \cos(   \beta_{\bk} ), \notag
\end{align}
which is real. In particular,  $\psi_{\bk}^{\ell m n}(\bJ) $ vanishes for odd values of $\ell+m+k_v$.
We detail this computation in Appendix~\ref{app:response_matrix}.

 One can check the validity of this expression by computing the potential basis elements via their angular Fourier decomposition \begin{align}
 F^{\ell m n}(\xi)\, Y_{\ell}^m(v,\phi) = \hspace*{-3mm} \sum_{\substack{k_u,k_v \\ \ell+m+k_v\, \mathrm{even} \\ k_{\phi}=m}} \hspace*{-4mm} W_{k_u k_v }^{\ell m n}(\bJ)\, e^{\ri \bk \cdot \btheta}.
 \end{align}

\subsection{Linear response of oblate spheroidal clusters}
The matrix elements given in equation~\eqref{eq:Mpq_generic} take the form \begin{equation}
\rM_{pq}\left(\omega\right)\!=\!\delta_{m^p}^{m^q}\frac{ 32\pi^4 G }{\Delta^2} \hspace*{-6mm} \sum_{\substack{k_u,k_v \\ \ell^p+m^p+k_v\, \mathrm{even}\\ \ell^q+m^q+k_v\, \mathrm{even}}} \hspace*{-6mm} \!\!\int \!\!\rd \bJ \frac{\bk \cdot \p F/\p \bJ}{\omega - \bk \cdot \bOmega }  W_{\bk}^{(p) }   W_{\bk}^{(q) } ,
\label{eq:Mpq_m} 
\end{equation}
where ${k_{\phi}\!=\!m^p\!=\!m^q}$. As a consequence, only matrix elements such that $\ell^p$ and $\ell^q$ have the same parity are non-zero. To compute the modes of the system, we must  solve the equation
\begin{equation}
\label{eq:matrix_frequency_equation}
\det[\bE(\omega)] = 0;\quad \, \bE(\omega) = \bI - \bM(\omega),
\end{equation}
where $\bE(\omega)$ is the dielectric matrix and $\bI$ the identity matrix. 

The system's axisymmetry, which appears through the Kronecker symbol  $\delta_{m^p}^{m^q}$, makes the response matrix a block-diagonal matrix. As such, one can decouple the system's instabilities through their azimuthal number, $m$, and study separately the axisymmetric modes $m=0$, the lop-sided modes $m=\pm 1$, the bi-symmetric modes $m=\pm 2$, and so on. Nonetheless, contrary to the infinitely thin  disc and the spherically symmetric sphere cases, the axisymmetric response matrix does not benefit from a two-dimensional reduction. As such, one has to face both theoretical and computational difficulties, such as the calculation of orbital frequency, of the angular Fourier transform of the basis elements, and the integration over the full three-dimensional action space.

\subsection{Azimuthal response matrix}
Owning to axial symmetry, we can decompose the dispersion relation  over the azimuthal numbers $m$ as
\begin{equation}
\det \bE(\omega) = \prod_m \det \bE^m(\omega),
\end{equation}
with $\bE^m(\omega)=\bI - \bM^m(\omega)$. Here, we define the azimuthal response matrix $\bM^m(\omega)$ by 
\begin{equation}
\rM_{pq}^m(\omega)\!=\! \frac{ 32\pi^4 G }{\Delta^2} \hspace*{-6mm} \sum_{\substack{k_u,k_v \\ \ell^p+m+k_v\, \mathrm{even}\\ \ell^q+m+k_v\, \mathrm{even}}} \hspace*{-4mm} \!\!\int \!\!\rd \bJ \frac{\bk \cdot \p F/\p \bJ}{\omega - \bk \cdot \bOmega }  W_{\bk}^{(p) }   W_{\bk}^{(q) } ,
\end{equation}
where ${\tbk=(k_u,k_v})$, ${k_{\phi}\!=\!m}$, ${p\!=\!(\ell^p,n^p)}$ and ${q\!=\!(\ell^q,n^q)}$. Axial symmetry yields the relation
\begin{equation}
\label{eq:mode_-m}
\rM_{pq}^{-m}(\omega)^{*}\!=\!\ \rM_{pq}^{m}(-\omega^{*}).
\end{equation}
Therefore, $\omega_0 + \ri \gamma$ is an $m$-mode if and only if $-\omega_0 + \ri \gamma$ is an $(-m)$-mode, and the analysis can be restricted  to  $m \geq 0$. 
In addition, for systems with DF depending only on $\Lz^2$ -- e.g., with no rotation -- the relation
\begin{equation}
\label{eq:mode_sym_omega}
\rM_{pq}^{m}(-\omega^{*})\!=\!\ \rM_{pq}^{m}(\omega)^{*},
\end{equation}
hold, which is obtained by making the change of variables ${\Lz \mapsto - \Lz}$.
We can thus limit our exploration to the sampling of frequency space over ${\rm{Re}(\omega) \!\geq\! 0}$.\footnote{These relations only hold \textit{a priori} on the upper frequency plane ${\mathrm{Im}(\omega) \!>\! 0}$, where the integral definition of the response matrix holds. However, they can be extended  to the lower part of complex plane after analytic continuation of the response matrix \citep[see, e.g.][ and Appendix~\ref{app:analytic_continuation}]{Weinberg1994,Fouvry2022}.}
In addition, following \citet{Rozier2022}, we can show that an accurate computation of the modes $m=-1,0,1$ requires the use of inertial pseudo-forces. For the sake of simplicity, we shall restrict our analysis to the bi-symmetric modes $m=2$.

\subsection{Mode shapes}

Finally, linear response theory enables us to represent the shape of the unstable modes using the eigenvectors of the response matrix. Specifically, when an unstable mode arises, we obtain an eigenvector, $\mathbf{a}=(a^m_{\ell n})$, such that 
\begin{equation}
\bM (\omega)\, \mathbf{a} =  \mathbf{a}.
\end{equation}
It follows that the shape of the density perturbation, $\delta \rho$, and the potential perturbation, $\delta \psi$, are given by the bi-orthogonal expansions
\begin{subequations}
\begin{align}
\delta \rho^m(\br) &= \sum_{\ell n} a^m_{\ell n} \,\frac{D^{\ell m n}(\xi)}{ \,c_{\Delta}(\xi,\eta)} \, Y_{\ell}^{m}(v,\phi), \\
\delta \psi^m(\br) &= \sum_{\ell n} a^m_{\ell n} \,F^{\ell m n}(\xi) \, Y_{\ell}^{m}(v,\phi).
\end{align}
\end{subequations}

\section{Application: Kuzmin--Kutuzov model}
\label{sec:KK_response}

\subsection{Description of the model}

The Kuzmin--Kutuzov model is an oblate \stak\ cluster. Indeed, its potential can be expressed in prolate spheroidal coordinates (see Section~\ref{sec:prolate_spheroidal} and Appendix~\ref{app:coordinates}) using equation~\eqref{eq:psi_stackel}, with the choice \citepalias[see, e.g., equation~{4.1} of][]{Dejonghe1988}
\begin{equation}
f(\tau)=G M (c-\sqrt{\tau}).
\end{equation}
It follows that its potential reads
\begin{equation}
\psi(\lambda,\nu)=-\frac{GM}{\sqrt{\lambda}+\sqrt{\nu}},
\end{equation}
while its associated mass density reads
\begin{equation}
\rho(\lambda,\nu) = \frac{M c^2}{4\pi} \frac{\lambda \nu + a^2(\lambda + 3 \sqrt{\lambda \nu} + \nu)}{(\lambda \nu)^{3/2}(\sqrt{\lambda}+\sqrt{\nu})^3}.
\end{equation}
Now, let use define the dimensionless quantities \citepalias[see table~1 of][]{Dejonghe1988}
\begin{subequations}
\label{eq:reduced}
\begin{align}
\tE &= -\frac{(a+c)E}{GM};\hspace{1mm}
\tLz= \frac{\Lz}{\sqrt{(a+c)GM}},\\
\ta& = \frac{a}{a+c};\hspace{1mm}
\tc = \frac{c}{a+c}.
\end{align}
\end{subequations}
Defining the squared eccentricity $h=1-(c/a)^2$, we can obtain the corresponding two-integral distribution function, $F(E,\Lz)$, under the form (\citetalias{Dejonghe1988}; \citealt{Batsleer1993})
\begin{align}
\label{eq:Ftot_KK}
F(E,\Lz)&\!=\!f_0 \tE^{5/2} \!\sum_{\epsilon=\pm 1} \!\int_0^1\!\! \frac{ \rd t(1-t^2)}{(1-2 \ta \tE t \sqrt{1-t^2} + \epsilon \sqrt{z} t)^5}\notag\\
&\times [(3\!+\!4 x_{\epsilon}\!-\!x_{\epsilon}^2)(1\!-\!x_{\epsilon})(1-t^2)\!+\!12t^2] ,\! 
\end{align}
where 
\begin{subequations}
\begin{align}
f_0&=\frac{M}{\big(GM(a+c)\big)^{3/2}}\frac{\tc^2}{2^{3/2}\pi^3 \ta},\\
z&=2 h \tE \tLz^2, \label{eq:def_z} \\
x_{\epsilon}&=2 \ta \tE t \frac{\sqrt{1-t^2}}{1+\epsilon\sqrt{z} t}.
\end{align}
\end{subequations}
Finally, one can verify that this distribution function reduces to the isotropic isochrone distribution \citep[see, e.g.,][]{Henon1960,Fouvry2021} in the spherical limit, and to the fully tangentially anisotropic Toomre distribution function \citep{Miyamoto1971} in the flat limit (see Appendix~\ref{app:isochrone_to_toomre} for more details). 

In this paper, we shall use the physical units defined by setting $G=M=a+c=1$.

\subsection{Non-rotating bi-symmetric modes}
\label{subsec:non_rotating_m_2_modes}

The bi-symmetric instabilities are identified  by solving Eq.~\eqref{eq:matrix_frequency_equation}. This is done by computing $\det \bE(\omega)$ on a grid in the upper complex frequency plane and searching for its zeros.

\begin{figure}
    \centering
     \includegraphics[width=0.45 \textwidth]{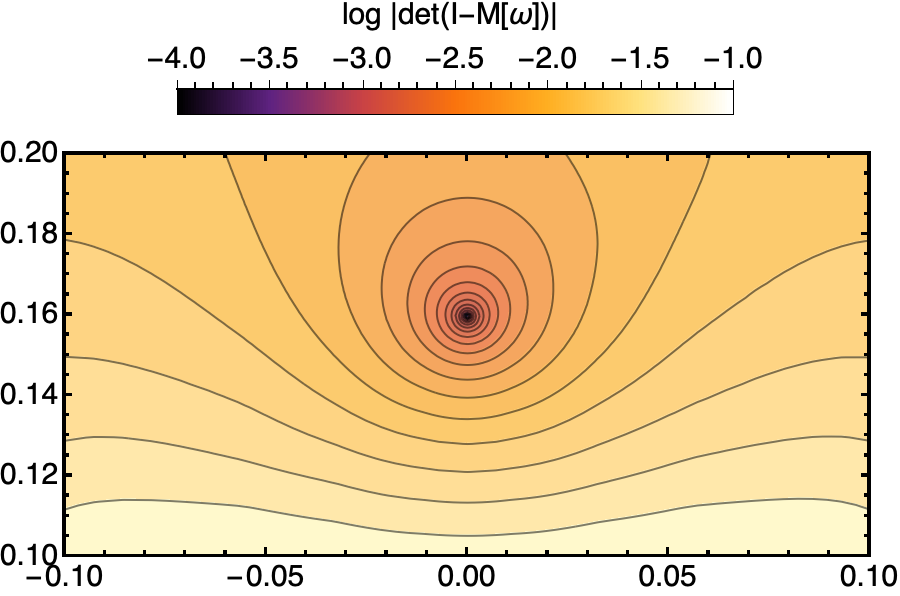}
   \caption{Isocontours of  $\det \bE(\omega)$ in the upper plane of the frequency space for ${m\!=\!2}$ for the clusters with ${a\!=\!0.88}$. We used the values ${\ell_{\max}\!=\!30}$, ${n_{\max}\!=\!10}$ and ${k_{\max}\!=\!10}$, $256\times256\times128$ sampling nodes for the $(J_u,J_v,\Lz)$  action integrals and 100 sampling nodes for the $W_{\bk}^{(p)}$ integrals. Here, we detect a growing mode at ${\omega \!=\!0.16 \, \ri}$.
   }
   \label{fig:modes_m_2}
\end{figure}

 This process is illustrated in Fig.~\ref{fig:modes_m_2} for the cluster of flatness $c/a=0.136$.
As increasingly flattened clusters are considered, a greater number of $\ell$ harmonics are required to achieve convergence toward the physical mode. This arises from the attempt to represent a highly flattened object using spherical harmonics, which are inherently better suited for small perturbations around a spherical shape.

\begin{figure}
    \centering
     \includegraphics[width=0.45 \textwidth]{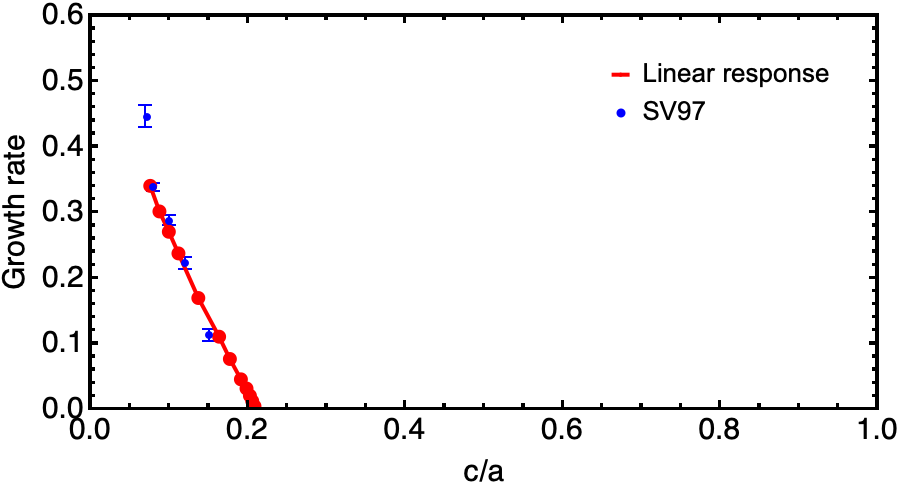}
   \caption{Growth rate as a function of flattening $c/a$, for the modes $m=2$ of non-rotating clusters. The predicted growth rate from linear response theory, in red, closely match those measured by  \citetalias{Sellwood1997}, shown in blue.
    }
   \label{fig:GR_m_2}
\end{figure}

Let us now investigate the impact of flattening on the stability of these clusters by applying this method to a wide range of flattening parameters, $c/a$.
Fig.~\ref{fig:GR_m_2} illustrates both an application of linear theory and its effectiveness in reproducing measurements from simulations.  
We obtain a 10\% agreement with the measurements of the bending modes made in $N$-body simulations by \citetalias{Sellwood1997}. Furthermore, we recover the same transition to stability at the threshold $c/a=0.208$.

\begin{figure}
    \centering
     \includegraphics[width=0.4 \textwidth]{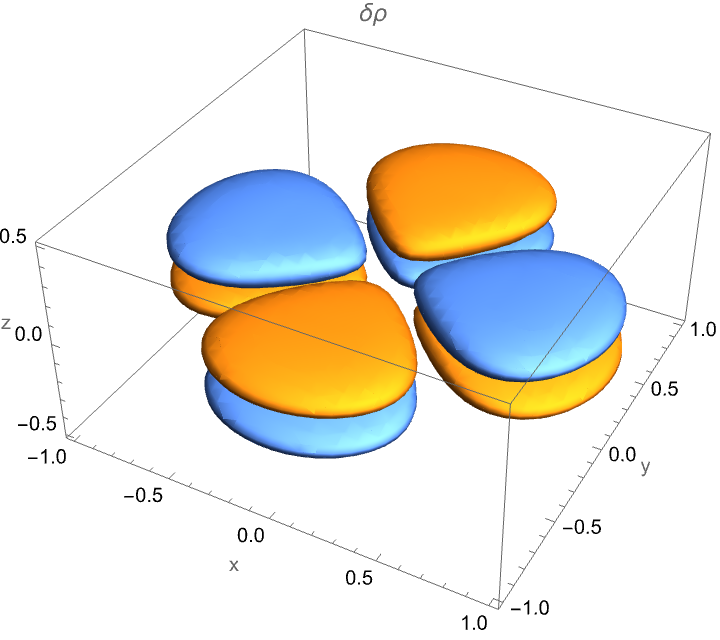}
   \caption{Shape of the ${m\!=\!2}$ density mode of a non-rotating, ${a\!=\!0.9}$ cluster, corresponding to a frequency ${\omega\!\sim\!0.22 \, \ri}$. 
   It is saddle-shaped, which corresponds to the shape of bending modes made by \citetalias{Sellwood1997}. 
   }
   \label{fig:mode_shape_m_2}
\end{figure}

Figure~\ref{fig:mode_shape_m_2} shows the shape of the density perturbation for the $a=0.9$ cluster, which appears as an saddle-shaped overdensity. 

\section{Introducing rotation} \label{sec:rotation}

\subsection{Response matrix with rotation}
Let us introduce rotation to the system by adding an odd component to the distribution function, following  Lynden-Bell's daemon prescription \citep{LyndenBell1960}
\begin{equation}
\label{eq:KKLB}
\Frot (E,\Lz) = F(E,\Lz) (1 + \alphar\, \sgn [\Lz]),
\end{equation}
where $\Frot (E,\Lz)$ is the DF of the rotating cluster, and $F(E,\Lz)$ the distribution function of the  non-rotating Kuzmin--Kutuzov cluster.  This choice is of course somewhat artificial, 
but convenient both for theoretical and numerical purposes. 
Then, the rotational response matrix may be decomposed into two contributions
\begin{equation}
\label{eq:Mpq_rot_alpha}
\rM_{pq}^m(\omega) = \rM_{pq}^{m,a}[\omega] +\alphar \big( \rM_{pq}^{m,b}[\omega]+ \rM_{pq}^{m,c}[\omega] \big),
\end{equation}
where we specify each component in Appendix~\ref{app:rotation}. 
We refer to Appendix~\ref{app:LBD_disc} for a discussion on the impact of the discontinuity of $\Frot$ on the modes' location.

\subsection{Spin parameter}

Let us define the spin parameter \citep{Peebles1969,Emsellem2007} \begin{equation}
\lambda_{\rr} = \frac{|\Lz^{\mathrm{tot}}| \sqrt{|E^{\mathrm{tot}}|}}{G M},
\end{equation}
where the specific mean energy and mean angular momentum are given by
\begin{subequations}
\begin{align}
 E^{\mathrm{tot}} &= \frac{(2\pi)^3}{M} \int \rd \bJ \, E(\bJ)\, \Ftot(\bJ),\\
 \Lz^{\mathrm{tot}}&=  \frac{(2\pi)^3\alphar}{M}  \int \rd \bJ \, |\Lz|\, \Ftot(\bJ).
\end{align}
\end{subequations}
Here, $\Ftot(\bJ)$ is the non-rotational Kuzmin--Kutuzov distribution function given by equation~\eqref{eq:Ftot_KK}, while $E$ and $\Lz$ are the classical specific integrals of motion. 
We note that $\lambda_{\rr}$ is proportional to $\alphar$ for a given value of $c/a$, which allows us to easily convert our observation for the Kuzmin cluster into its
classical  $\lambda_{\rr}$ formulation \citep{Emsellem2007}.

\subsection{Rotating bi-symmetric Kuzmin--Kutuzov clusters}

\begin{figure}
    \centering
           \includegraphics[width=0.45 \textwidth]{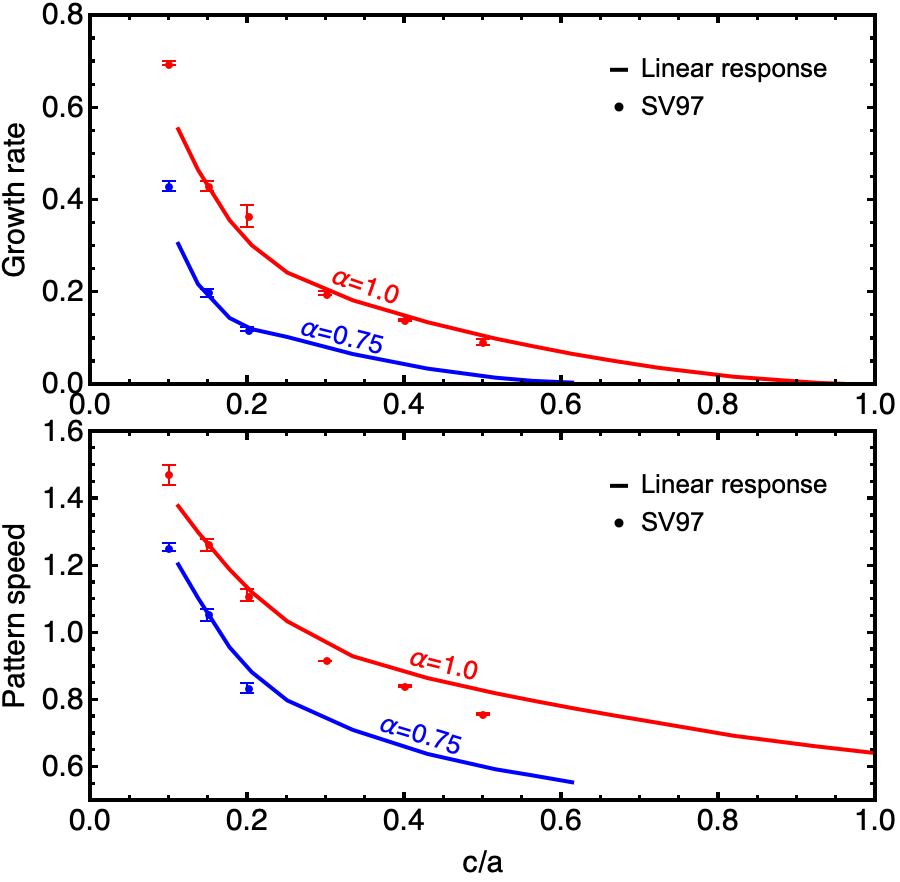}
\caption{Growth rate (top panel) and pattern speed (bottom panel) as a function of flattening $c/a$, for the modes $m=2$ of fast rotating clusters $\alphar=0.75$ (in blue) and maximally rotating clusters $\alphar=1$ (in red). Full lines are the predicted growth rate from linear response theory, while dots are those measured by  \citetalias{Sellwood1997}. The stability of the Kuzmin--Kutuzov cluster increases as it approaches the spherical limit, with the maximally rotating configuration being the last to achieve stability at   $c/a=1$. }
   \label{fig:GR_v_flattening_fixed_rotation}
\end{figure}

In the same spirit as in Section~\ref{subsec:non_rotating_m_2_modes}, let us compare the prediction from linear response theory to measurements made in $N$-simulations by \citetalias{Sellwood1997}. The impact of flattening on growth rate for a rapidly rotating cluster $\alpha_r=0.75$ and the maximally rotating cluster $\alpha_r=1$ is shown in Fig.~\ref{fig:GR_v_flattening_fixed_rotation}. 
First, we note an agreement with $N$-body measurements within 10\% accuracy, as well as matching behavior w.r.t. flattening ratio. In addition, we observe a steep increase of the growth rate as one considers very flattened clusters. Conversely, flattened systems tend to become more and more stable as they become rounder for any rotation parameters. In particular, the cluster with rotation parameter $\alpha_r=0.75$ become completely stable at $c/a \sim 0.6$, while the maximally rotating clusters become stable at $c/a=1$.\footnote{This is reminiscent of conclusions made by \cite{Rozier2019}, who showed that the similar isotropic Plummer cluster was stable for any rotation parameter.}  While \citetalias{Sellwood1997} appear not to measure any such instabilities beyond $c/a=0.6$, this may be explained either by the low growth rate prediction by linear theory ($\gamma <0.07$), and/or by the impact of softening on the growth rate \citep{Rijcke2019,Roule2024}.
Finally, linear theory provides us with a tool to systematically perform the stability analysis w.r.t. to the flattening ratio of the cluster, $c/a$, and its rotation parameter, $\alphar$, which we convert into the associated spin parameter, $\lambda_{\rr}$. 
\begin{figure}
    \centering
   \includegraphics[width=0.45 \textwidth]{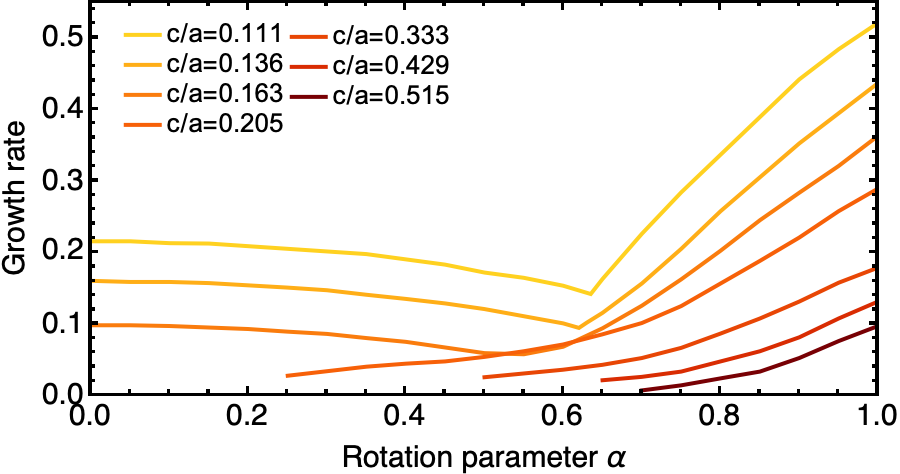}
   \caption{
   The growth rate of the dominant 
 $m=2$ mode is shown as a function of rotation for clusters with a fixed flattening ratio 
$c/a$. A clear transition is observed from bending modes to bar-growing modes in highly flattened clusters, with the growth rate reaching a minimum at the point of transition. This is illustrated in Fig.~\ref{fig:modes_subdominant} of Appendix~\ref{app:bend_v_bar}, which displays the  behavior of each modes in their subdominant regimes.
 }
   \label{fig:GR_ratio_slices}
\end{figure}
\begin{figure*}
    \centering
\includegraphics[width=0.9 \textwidth]{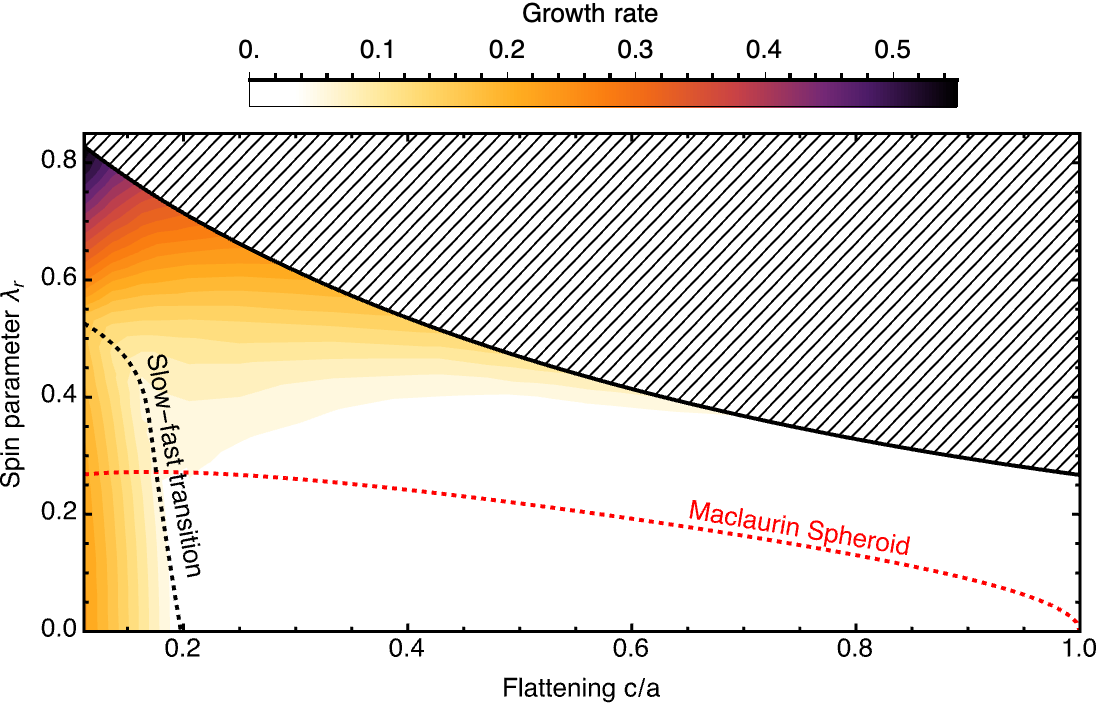}
   \caption{Dependency of the $m=2$  growth rates w.r.t. flattening $c/a$ and spin parameter, $\lambda_{\rr}$. The hashed region corresponds to the parameters which cannot be attained by a single-component St\"ackel cluster. For highly flattened clusters, a transition (dashed black line) between slowly rotating bending modes (Fig.~\ref{fig:mode_shape_rot_m_2_slow}, left panel) and rapidly rotating bar-growing modes (Fig.~\ref{fig:mode_shape_rot_m_2_slow}, right panel) occurs. For models rounders than $c/a=0.2$, the clusters are either stable or subject to the latter instability. We refer to Fig.~\ref{fig:GR_ratio_slices} for a sample of $c/a$-slices of the growth rate. Finally, we show in bashed red the behavior of the stable Maclaurin spheroids for comparison purpose, which display a much different behavior than the Kuzmin--Kutuzov clusters. }
   \label{fig:GR_flattening_rotation}
\end{figure*}

\begin{figure*}
    \centering
\includegraphics[width=0.8 \textwidth]{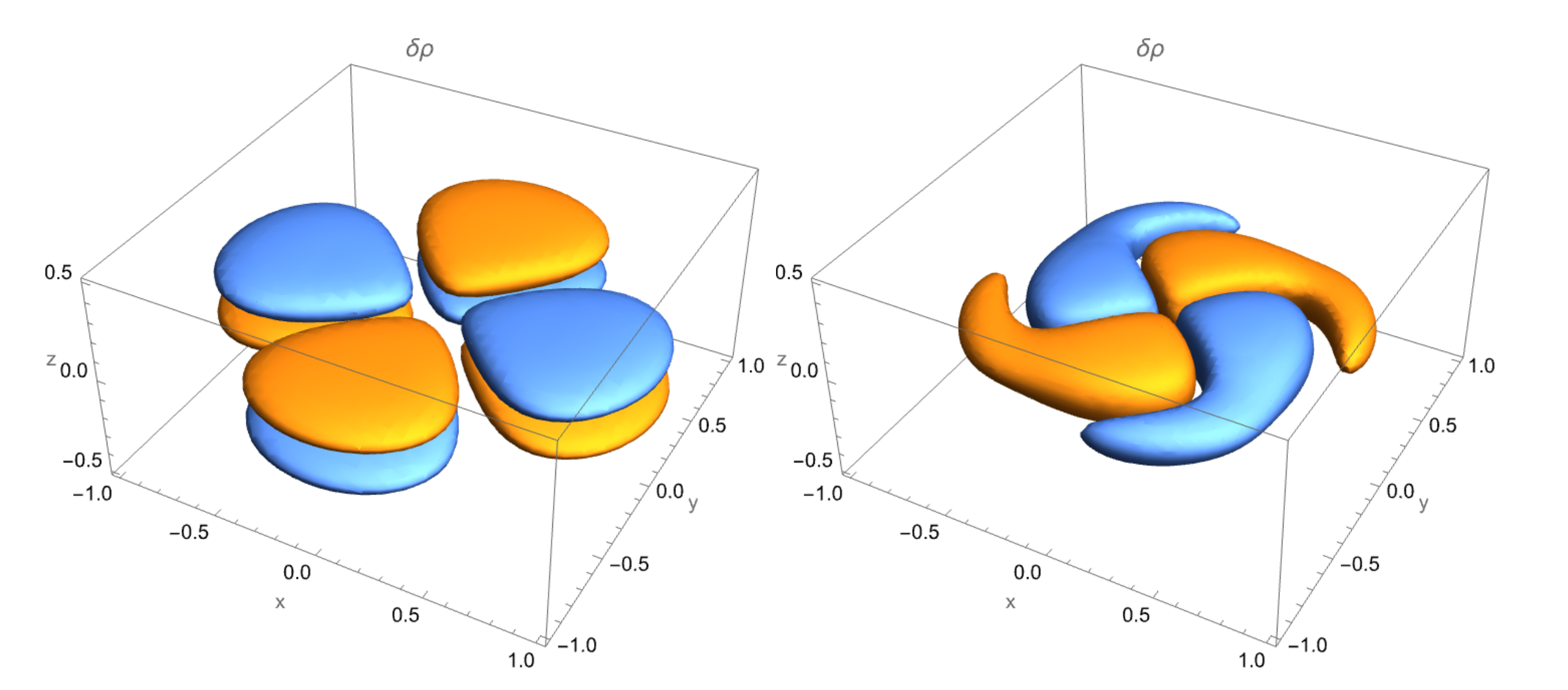}
   \caption{Shape of the $m=2$ density modes of a $a=0.9$ cluster, for the two types of rotating instabilities. {\sl Left panel:} slow bending mode ($\alphar=0.4$, $\omega=0.12 + 0.19 \,\ri$). {\sl Right panel:} Fast bar-growing mode ($\alphar=1.0$, $\omega=1.31 + 0.53\, \ri$). We highlight the impact of rotation in  Fig.~\ref{fig:shape_bending_rot_slice} of appendix~\ref{app:comp_fig_bend_mode_shape}. The fast rotating mode, however, corresponds to a distinct family of instability, with a much higher pattern speed, and strikingly, presents a very different shape. 
}
   \label{fig:mode_shape_rot_m_2_slow}
\end{figure*}

We represent this overall analysis in  Figs.~\ref{fig:GR_ratio_slices} and~\ref{fig:GR_flattening_rotation}. We identify four regions of interest. First, we recover the dependence of the stability of non-rotating clusters on their flattening by examining the $\lambda_{\rr}=0$ axis, where the stability threshold occurs at $c/a=0.208$. Secondly, in highly flattened systems with relatively low spin parameters, the growth rate decreases as rotation increases. These are the bending modes measured by \citetalias{Sellwood1997}, and are illustrated in the left panel of Fig.~\ref{fig:mode_shape_rot_m_2_slow}. However, beyond a given rotation threshold, which depends on the cluster's flattening, the dominant mode is a so-called bar-growing mode, whose shape presents  a distinctive bi-symmetric spiral structure (see right panel of  Fig.~\ref{fig:mode_shape_rot_m_2_slow}). Moreover, these modes seem to persist across a much broader range of flattening ratios, with the maximally rotating clusters being the last to stabilize as the system approaches the spherical limit.
Finally, there appears to be a region of stability in the parameter space for cluster which are not too flattened and have a low enough rotation parameter.

\subsection{Transition to damped bar modes}

\begin{figure}
    \centering
     \includegraphics[width=0.45 \textwidth]{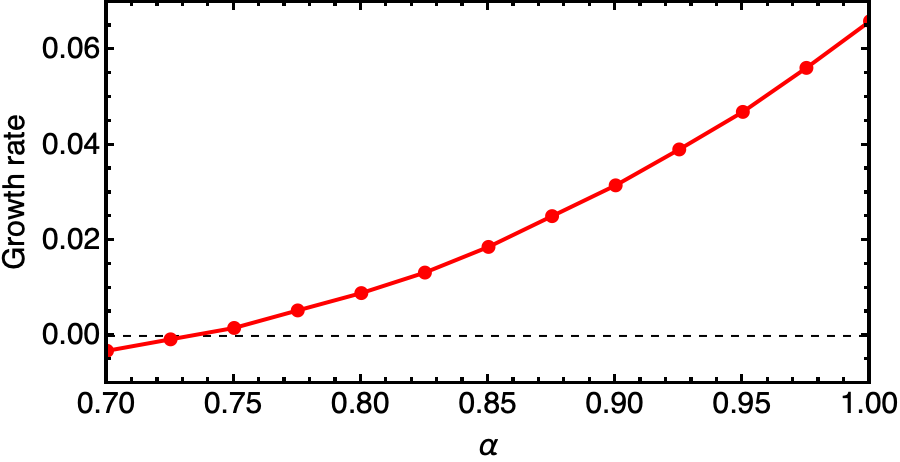}
   \caption{Evolution of the growth rate of a $c/a=0.613$ cluster w.r.t. the rotation parameter $\alphar$. Transition from damped bar modes (on the left) to unstable bar modes occurs at a fixed value as we increases $\alphar$ .  }
   \label{fig:GR_damped_bar_m_2}
\end{figure}

We may perform analytic continuation on the matrix elements in order to study the transition to stability of the bar-growing modes. We refer to Appendix~\ref{app:analytic_continuation} for further details.
Figure~\ref{fig:GR_damped_bar_m_2} shows the growth rate of a $c/a=0.613$ cluster w.r.t. the rotation parameter, $\alphar$.
The cluster becomes stable at a non-zero rotation parameter, before exhibiting damped modes below that threshold.

However, such stability analysis cannot be performed for an arbitrary cluster. Indeed, the calculation of the  response matrix  close to the real frequency axis may become a numerical challenge due the resonant denominator -- especially for low values of $\mathrm{Re} (\omega)$. 
Therefore, when studying the marginal stability of non-rotating clusters around 
$c/a=0.208$, careful attention must be given to the integration in action space. While a change of variables similar to that used by \citet{Fouvry2022} might be of interest, such a transformation has yet to be developed within the context of axisymmetric systems.

\section{Conclusions and perspectives}
\label{sec:discussion}
Nearly 30 years after the seminal work of \cite{1995PhDT.......119R}, this paper extended linear response theory to rotating oblate  \stak\, clusters. Our implementation was validated against the results from \citetalias{Sellwood1997}, showing that the converged  $m=2 $ modes of flattened and rotating spheroids closely match $N$-body simulations at the $\sim$10\% level in terms of growth rates (see Figs.~\ref{fig:GR_m_2} and ~\ref{fig:GR_v_flattening_fixed_rotation}), pattern speeds, and shapes. When rotation is introduced, two families of fast and slow modes are found (Figs.~\ref{fig:mode_shape_m_2} and~\ref{fig:mode_shape_rot_m_2_slow}): the bending modes, which are saddle-shaped, and the bar-growing modes, which are spiral shaped. Our implementation also allows for analytic continuation so that certain weakly damped modes can be identified (see Fig.~\ref{fig:GR_damped_bar_m_2}). However, a more systematic study (for, e.g., the marginal stability of non-rotating clusters) would require a more careful treatment of the resonant denominator of the response matrix, which becomes sharp near the real frequency line.

Note that breaking spherical symmetry introduces various theoretical and computational challenges when applying linear response theory. While axial symmetry allowed us  to  decompose the response matrix into distinct azimuthal components, it also fundamentally transformed the computational problem into a three-dimensional one.  Specifically, computing a $5 \!\times\! 5$ frequency-space grid for the response matrix of a single cluster with $(\ell_{\max},n_{\max},k_{\max})=(30,10,10)$, and performing the action-space integration over $256\!\times\!256\!\times\!128$ nodes, required approximately 38 hours across 512 CPU cores.  Hence we had to carefully balance mode convergence with computational cost to achieve accurate predictions.
On the other hand, this formalism provides access to the system's linear response across all values of $\alpha_r$,  as per equation~\eqref{eq:Mpq_rot_alpha}. Consequently, future studies may find it advantageous to revisit this type of analysis with improved computational resources.

Although this single-component asymptotically cold model is not realistic enough to compare against data \citep[e.g.][and subsequent work]{Emsellem2007} and simulations \citep[e.g.][]{Choi2018}, the  formalism presented here may easily be extended to multi-component St\"ackel systems, and therefore 
eventually be applied to  more realistic equilibria.

\subsection{Perspectives}

One should first investigate the limitations of the specificities of this  \stak\ model, which extends the one-component, isotropic isochrone sphere to flattened systems. 
Further exploration would be beneficial to extend the computation of linear response to more realistic spheroids, incorporating three-integral distribution functions \citep[see, e.g., section~{IV.d} of][]{Dejonghe1988}, two- or three-component  \stak\, discs and halos \citep{Dejonghe1988b,Hunter1993,Batsleer1994,Famaey2003,Petac2019,Gromov2021,Koppelman2021}, or empirically fitted distributions \citep{Dejonghe1989,Famaey2002}. It would also be interesting to examine the linear stability of elliptical clusters \citep{1982MNRAS.199..171W,1991ARA&A..29..239D,1999PASP..111..129M,2008MNRAS.388.1321P} or to investigate more sophisticated models of rotation, though computational complexity might quickly become a heavily limiting factor.
Furthermore, the transposition the present formalism into the time domain following, e.g., \citet{Rozier2022} (and references therein), remains an open problem, as is the inclusion of pseudo-forces in the calculation of axisymmetric and lop-sided instabilities.

Beyond linear response, the methodology could be naturally expanded to kinetic theory \citep{Roule2022,Roule2024}, say, to address the secular vertical heating of thick discs while improving upon  perturbative studies \citep{2017MNRAS.471.2642F}. Such investigations could offer deeper insights into the evolutionary trajectories of galactic systems, enabling comparisons across galaxy populations over cosmic timescales and within diverse environments \citep{2006SSRv..123..485G, 2022ApJ...938L..15C, 2022ApJ...940L..14N}.

Eventually, understanding statistically\protect~\citep[beyond the mean response, see][]{Touchette2009,Feliachi+2023}   the impact of rotation and flattening 
should prove critical to model the morphological evolution of populations of galaxies across cosmic times~\protect\citep[see, e.g.\@,][]{10.1111/j.1365-2966.2008.13689.x,Conselice2014}.
Indeed, the distribution of angular momentum is, beyond  mass,
the most relevant dynamical parameter for such populations:
it separates elliptical galaxies from spiral ones \citep{10.1111/j.1365-2966.2006.10115.x}.
The long-term impact of rotation
on galactic orbital structures will therefore be the focus of our future research.

\section*{Data Distribution}
The  data underlying this article 
is available through reasonable request to the author.
The notebook used to compute the bi-orthogonal basis function is  distributed at the URL: \href{https://github.com/KerwannTEP/SPOCK}{https://github.com/KerwannTEP/SPOCK}. 
The code used for the computation of the linear response, written in \texttt{Julia} \citep{JuliaCite}, is also now part of the general purpose  \texttt{Julia} stellar dynamics codes found at  \href{https://github.com/JuliaStellarDynamics}{https://github.com/JuliaStellarDynamics}, as discussed by  \cite{Petersen2024}.

\section*{Acknowledgements}
We are grateful to J.-B. Fouvry, M. Roule, M. Weinberg and A.~L.~Varri 
for numerous suggestions during the completion of this work,
which is partially supported by the grant \href{https://www.secular-evolution.org}{\emph{SEGAL}} ANR-19-CE31-0017
of the French Agence Nationale de la Recherche and by the National Science Foundation under Grants No. AST-2310362 to the University of North Carolina and No. PHY-2309135 to the Kavli Institute for Theoretical Physics (KITP). 
We thank St\'ephane Rouberol for the smooth running of the
Infinity cluster, where part of the computations was performed. We would like to also thank the University of North Carolina at Chapel Hill and the Research Computing group for providing computational resources and support that have contributed to these research results.

\appendix

\counterwithin{figure}{section}
\renewcommand\thefigure{\thesection\arabic{figure}}

\section{Prolate spheroidal coordinates}
\label{app:coordinates}

\subsection{Definition and conventions}
\label{app:convention_spheroidal}

\begin{figure*}
    \centering
     \includegraphics[width=0.8 \textwidth]{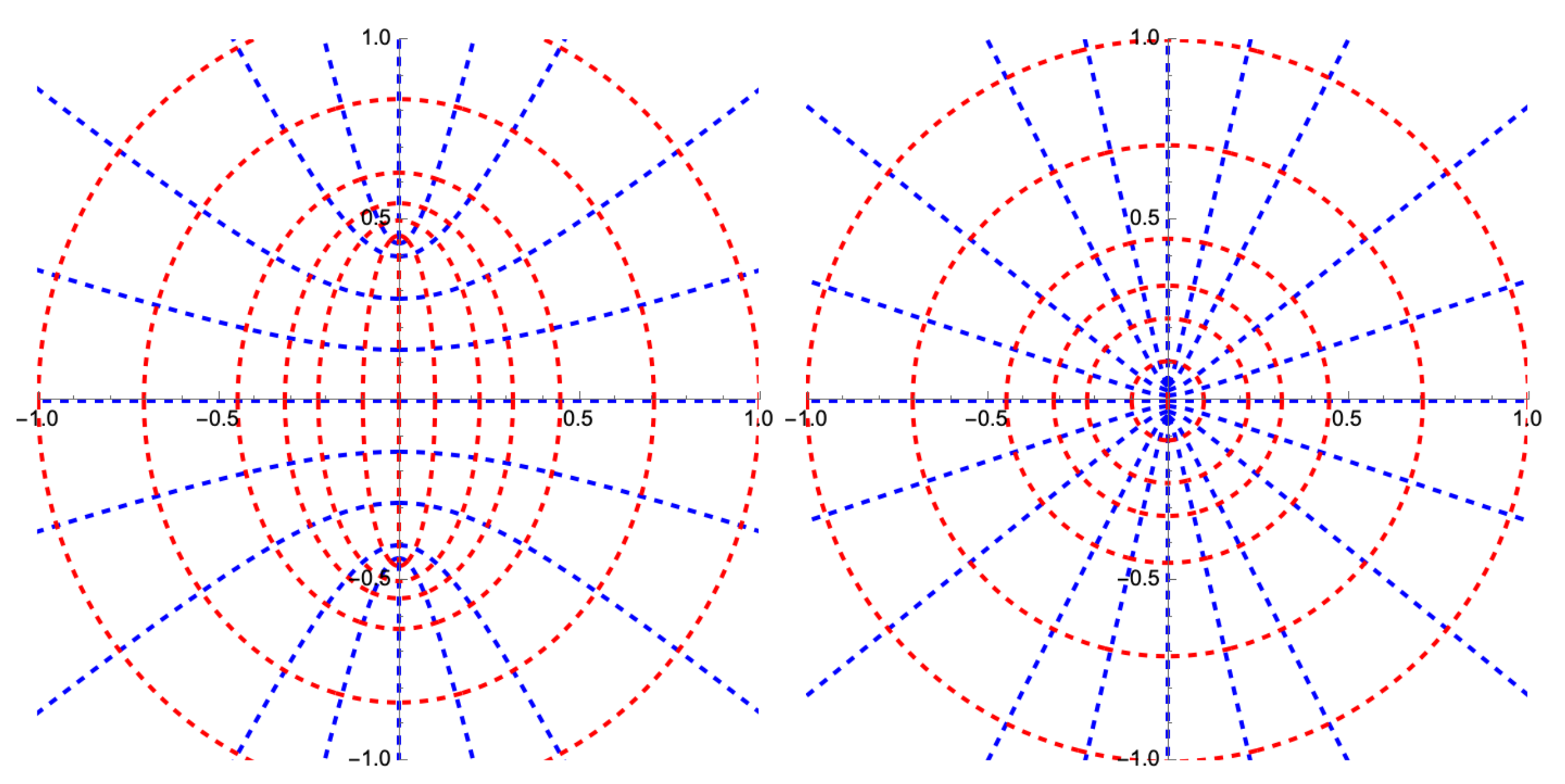}
\caption{Two-dimensional cut of the spheroidal coordinate system, viewed sideway to the equatorial plane. Blue dashed lines are contours of constant $\eta$, while red dashed lines are contours of constant $\xi$. On the left panel, $a=0.6$ and $c=0.4$. The two focal points are located at $z=\pm \Delta$, with $\Delta=0.447$ in this figure. On the right panel, $a=0.501$ and $c=0.499$. The spheroidal coordinates converge to the spherical coordinates in the spherical limit. }
   \label{fig:LambdaNu}
\end{figure*}

We defined in Section~\ref{subsec:prolate_coordinates} the \textit{confocal elliptic coordinates} $(\lambda,\nu,\phi)$, which  straightforwardly relate to cylindrical coordinates
as 
\begin{equation}
R^2=\frac{(\lambda-a^2)(\nu-a^2)}{c^2-a^2};\quad z^2=\frac{(\lambda-c^2)(\nu-c^2)}{a^2-c^2}.
\end{equation}
We illustrate this coordinate system in Fig.~\ref{fig:LambdaNu}.
While this formulation is widely used in old literature, more recent papers tend to use the \textit{elliptic coordinates} $(u,v)$ \citep[see, e.g.,][]{Binney2012}, which parametrize the \textit{confocal elliptic coordinates} as
\begin{subequations}
\label{eq:elliptic_coordinates}
\begin{align}
\lambda &= a^2 \cosh^2 u -c^2 \sinh^2 u = c^2 + \Delta^2 \cosh^2 u,\\
\nu &= c^2 \sin^2 v +a^2 \cos^2 v= c^2 +\Delta^2 \cos^2 v.
\end{align}
\end{subequations}
In particular, it follows that
\begin{equation}
R = \Delta \sqrt{(\xi^2-1)(1-\eta^2)};\quad z = \Delta\, \xi\, \eta,
\end{equation}
where we define the coordinates $(\xi,\eta)$ by setting
\begin{equation}
\label{eq:xi_eta}
\xi = \cosh u;\quad \eta = \cos v,
\end{equation}
as they are useful when using potential-density basis elements \citep{Robijn1996b}. Finally, the reverse transformation $(R,z)\mapsto (\xi,\eta)$ is given by
\begin{subequations}
\begin{align}
\xi  &= \frac{1}{2\Delta} \big(\sqrt{R^2+(z+\Delta)^2} + \sqrt{R^2+(z-\Delta)^2}\big),\\
\eta  &= \frac{1}{2\Delta} \big(\sqrt{R^2+(z+\Delta)^2} - \sqrt{R^2+(z-\Delta)^2}\big).
\end{align}
\end{subequations}

\subsection{Relations between $(u,v,\phi)$  and $(\lambda,\nu,\phi)$ coordinates}
\label{app:Ju_to_Jlambda}

If we denote by $I_3$ the third integral in $(u,v,\phi)$ coordinates and by $\widetilde{I}_3$ that in $(\lambda,\nu,\phi)$ \citepalias[see, e.g.,][]{Dejonghe1988}, then we have the relation
\begin{equation}
\widetilde{I}_3 = \frac{1}{2} \big( 2\Delta^2 E + 2 \Delta^2 I_3 - \Lz^2 \big),
\end{equation}
for any axisymmetric St{\"a}ckel potential, which we can obtain by evaluating $p_u$ and $p_{\lambda}$ at their boundaries of motions, coupled with equations~\eqref{eq:elliptic_coordinates}.
In particular, letting $I_2=\Lz^2/2$, we obtain the spherical limit
\begin{equation}
 I_2 + \widetilde{I}_3 \rightarrow \frac{L^2}{2}.
\end{equation}
A large fraction of distribution functions referenced in the literature (\citetalias[see, e.g.,][]{Dejonghe1988}; \citealt{Robijn1996,Famaey2002, Famaey2003}) involving a third integral, should be understood as functions of $(E,I_2,\widetilde{I}_3)$. They can be converted to functions of $(E,I_2,I_3)$ by using the above equation.

Now, we can show from this relation that
\begin{equation}
 p_{\lambda}^2 = 4 \Delta^2 \cosh^2 u \sinh^2 u \,p_u^2 ,
\end{equation}
where $p_{\lambda}$ is the conjugate momentum to $\lambda$ defined by \citepalias{Dejonghe1988}
\begin{equation}
p_{\lambda}^2 = \frac{1}{2(\lambda-a^2)} \bigg(E - \frac{I_2}{\lambda-a^2} - \frac{\widetilde{I}_3}{\lambda-c^2} - \frac{f[\lambda]}{\lambda-c^2} \bigg).
\end{equation}
Therefore, for a given orbit, the action variables associated to $u$ and $\lambda$ are equal, i.e.
\begin{equation}
J_{\lambda} = \frac{1}{\pi} \int_{\lambda_0}^{\lambda_1}\hspace*{-2mm} \rd \lambda \,p_{\lambda} = \frac{1}{\pi} \int_{u_0}^{u_1}\hspace*{-2mm} \rd u\, p_{u}   = J_u.
\end{equation}
Similarly, one can show that
\begin{equation}
 p_{\nu}^2 = 4 \Delta^2 \cos^2 v \sin^2 v \,p_{v}^2 ,
\end{equation}
where $p_{\nu}$ is the conjugate momentum to $\nu$ defined by \citepalias{Dejonghe1988}
\begin{equation}
p_{\nu}^2 = \frac{1}{2(\nu-a^2)} \bigg(E - \frac{I_2}{\nu-a^2} - \frac{\widetilde{I}_3}{\nu-c^2} - \frac{f[\nu]}{\nu-c^2} \bigg),
\end{equation}
and therefore
\begin{equation}
J_{\nu} = \frac{2}{\pi} \int_{c^2}^{\nu_1}\hspace*{-2mm} \rd \nu \,p_{\nu} = \frac{2}{\pi} \int_{\frac{\pi}{2}}^{v_1}\hspace*{-2mm} \rd v\, p_{v}   = J_v.
\end{equation}

\section{Changes of coordinates}
\label{app:variables}

The transformation $(E,I_3,\Lz) \mapsto (J_u,J_v,\Lz)$ is carried out by a straight computation of the two action integrals. However, the inverse transformation lacks an explicit analytical expression. Nevertheless, it can be computed using Newton's method.

Let us define the vectors $\bJ=  (J_u,J_v,\Lz)^{\rT}$ and $\bw=(E,I_3,\Lz)^{\rT}$. Let use also define the function $\bfct(\bw)=(J_u[\bw],J_v[\bw],\Lz)^{\rT} - \bJ$. 
We wish to solve $\bfct(\bw)=0$. We start from an initial guess $\bw_0=(E_i,I_{3,i},L_z)$ with $I_{3,i}=\Lz^2/(2 \Delta^2) - E_0 $ and $E_i=E_s(\Lz,I_{3,i})/2$, which always describes a bound orbit for a Kuzmin--Kutuzov potential. Then, we define for each step
\begin{equation}
\bw_{n+1} = \bw_n - [J(\bw_n)]^{-1} \bfct(\bw),
\end{equation}
where $J(\bw)$ is the Jacobian given by
\begin{align}
\label{eq:Jacobian}
J(\bw)&= \begingroup
\renewcommand*{\arraystretch}{1.8}
 \begin{pmatrix}
 \displaystyle{\frac{\p J_u}{\p E}  }   &\displaystyle{\frac{\p J_u}{\p I_3}  } &\displaystyle{\frac{\p J_u}{\p \Lz}  }   \\
 \displaystyle{\frac{\p J_v}{\p E}  }   &\displaystyle{\frac{\p J_v}{\p I_3}  } &\displaystyle{\frac{\p J_v}{\p \Lz}  }  \\
0& 0&1  
\end{pmatrix}
\endgroup.
\end{align}
At each step, we should be careful not to go outside of the authorized region of $(E,\Lz,I_3)$. 

\section{St{\"a}ckel potential}
\label{app:Stackel_potential}

\subsection{Boundaries of motion}
\label{app:boundaries}

We may  define the effective potentials
\begin{subequations}
\begin{align}
U_{\eff}(u,I_3,\Lz)&=  \frac{\Lz^2 }{2 \Delta^2 \sinh^4 u}+\frac{ U(u)+I_3}{\sinh^2 u} ,\\
V_{\eff}(v,I_3,\Lz)&=  \frac{\Lz^2 }{2 \Delta^2 \sin^4 v}-\frac{  V(v)+I_3}{\sin^2 v} ,
\end{align}
\end{subequations}
such that 
\begin{subequations}
\begin{align}
p_u^2 &= 2 \Delta^2  \sinh^2 u\,( E  - U_{\eff}[u,I_3,\Lz]) ,\\
p_v^2 &= 2 \Delta^2\sin^2 v\,( E  - V_{\eff}[v,I_3,\Lz]).
\end{align}
\end{subequations}
The boundaries of motions are then computed by solving the equations ${p_u^2[u,I_3,\Lz]\!=\!0}$  and ${p_v^2[v,I_3,\Lz]\!=\!0}$ using  bisection.

For given $(I_3,\Lz)$, the shell orbits, i.e., the orbits of constant coordinates $u$, are given by finding the minimum of $  U_{\eff}[u,I_3,\Lz]$ w.r.t. $u$.

\subsection{Orbital frequencies and action derivatives}

Let us compute the orbital frequencies $\bOmega=(\Omega_{u},\Omega_{v},\Omega_{\phi})$. Following \citet{Binney2012}, the matrix
\begin{align}
\label{eq:freq_matrix}
\begingroup
\renewcommand*{\arraystretch}{1.5}
 \begin{pmatrix}
\Omega_u  & \Omega_v &\Omega_{\phi}   \\
 \displaystyle{\frac{\p I_3}{\p J_u}  }   &\displaystyle{\frac{\p I_3}{\p J_v}  } &\displaystyle{\frac{\p I_3}{\p \Lz}  }  \\
0& 0&1  
\end{pmatrix}
\endgroup
,
\end{align}
where $\Omega_t = \p E/\p J_t$ with $t=u,v,\phi$, 
is the inverse of the matrix
\begin{align}
\label{eq:grad_mat}
\begingroup
\renewcommand*{\arraystretch}{1.8}
 \begin{pmatrix}
 \displaystyle{\frac{\p J_u}{\p E}  }   &\displaystyle{\frac{\p J_u}{\p I_3}  } &\displaystyle{\frac{\p J_u}{\p \Lz}  }   \\
 \displaystyle{\frac{\p J_v}{\p E}  }   &\displaystyle{\frac{\p J_v}{\p I_3}  } &\displaystyle{\frac{\p J_v}{\p \Lz}  }  \\
0& 0&1  
\end{pmatrix}
\endgroup.
\end{align}
To compute the derivative in this second matrix, we need the derivatives of $p_u$ and $p_v$ with respect to $E$, $I_3$ and $\Lz$
\begin{subequations}
\label{eq:pupv_derivatives}
\begin{align}
\frac{\p p_u}{\p E} &=\frac{\Delta^2 \sinh^2 u}{p_u},&\frac{\p p_v}{\p E} &=\frac{\Delta^2 \sin^2 v}{p_v},\\
\frac{\p p_u}{\p I_3} &=-\frac{\Delta^2 }{p_u},&\frac{\p p_v}{\p I_3} &=\frac{\Delta^2 }{p_v},\\
\frac{\p p_u}{\p \Lz} &=-\frac{\Lz }{p_u \sinh^2 u},&\frac{\p p_v}{\p \Lz} &=-\frac{\Lz }{p_v \sin^2 v },
\end{align}
\end{subequations}

\subsection{Effective anomaly}
\label{app:reffective_anomaly}

Finding an explicit effective anomaly like in the Keplerian case  \citep{BarOr2016}, the isochrone case \citep{Fouvry2021} or the Plummer case  \citep{Tep2022} is a very difficult process, and requires a case-by-case exploration. In  general, an explicit analytical effective anomaly does not exist, and we must content with a numerical evaluation \citep{Henon1971,Roule2022,Petersen2024}. Let $t=u,v$. Following \citet{Eyre2010}, we define $\bar{t}=(t_0+t_1)/2$ and $\hat{t}=(t_1-t_0)/2$. We let 
\begin{equation}
t = \hat{t} \sin \theta + \bar{t},
\end{equation}
for $-\pi/2 \leq \theta \leq \pi/2$. In particular, 
\begin{equation}
 (t-t_0)(t_1-t) = \hat{t}^2 \cos^2 \theta.
 \end{equation}
Now, in order to compute frequencies and inverse coordinate transformations, we wish to compute integrals of the form
\begin{equation}
I=\int_{t_0}^{t_1} \rd t \frac{g(t)}{p_t(t)},
 \end{equation}
for some well defined function $g(t)$. We know that the momentum behaves as
\begin{equation}
 p_t(t) = A \sqrt{|t-t_x|} + \frac{B}{\sqrt{|t-t_x|}} + ...,
 \end{equation}
when $t \rightarrow t_x$, for $t_x=t_0,t_1$ and some constants $A$ and $B$. Therefore, we may write
\begin{equation}
 p_t(t) = \sqrt{(t-t_0)(t_1-t)}   \tilde{p}_t(t).
 \end{equation}
This makes $1/\tilde{p}_t(t)$ a theoretically perfectly well-defined function. It follows from a change of variable $t \mapsto \theta$ that
\begin{equation}
I=\int_{-\pi/2}^{\pi/2} \rd \theta  \frac{g(t[\theta])}{ \hat{p}_t(t[\theta])},\quad\hat{p}_t(t[ \theta]) = \frac{p_t(t[  \theta])}{\hat{t} \cos \theta},
 \end{equation}
making the integrand non-singular. We can then evaluate it usual standard quadrature techniques. The limit at the boundaries of motion $t=t_0,t_1$ are given by
\begin{equation}
   \tilde{p}_t(t) \rightarrow \frac{1}{\sqrt{t_1-t_0}} \sqrt{\bigg| \frac{\p (p_t^2)}{\p t}\bigg|}_{t=t_0,t_1}.
 \end{equation}

\subsection{Action derivatives of the momenta}

 In order to compute the angles, we need to compute the action derivatives of $p_u$ and $p_v$, which are to be understood as functions of $(J_u,J_v,\Lz)$. We get
\begin{subequations}
 \begin{align}
 \frac{\p p_t}{\p J_u} &=  \frac{\p p_t}{\p E} \frac{\p E}{\p J_u} +  \frac{\p p_t}{\p \Lz} \frac{\p \Lz}{\p J_u} +  \frac{\p p_t}{\p I_3} \frac{\p I_3}{\p J_u} ,\\
  \frac{\p p_t}{\p J_v} &=  \frac{\p p_t}{\p E} \frac{\p E}{\p J_v} +  \frac{\p p_t}{\p \Lz} \frac{\p \Lz}{\p J_v} +  \frac{\p p_t}{\p I_3} \frac{\p I_3}{\p J_v} ,\\
    \frac{\p p_t}{\p \Lz} &=  \frac{\p p_t}{\p E} \frac{\p E}{\p \Lz} +  \frac{\p p_t}{\p \Lz} \frac{\p \Lz}{\p \Lz} +  \frac{\p p_t}{\p I_3} \frac{\p I_3}{\p \Lz} ,
 \end{align}
 \end{subequations}
where $t=u,v$ and the momenta of the right hand-side are functions of $(E,\Lz,I_3)$. We may rewrite this system of equations in the matrix form
\begin{align}
\label{eq:relation_grads}
\begingroup
\renewcommand*{\arraystretch}{2}
\begin{pmatrix}
 \displaystyle{\frac{\p p_t}{\p J_u}  }  \\
 \displaystyle{\frac{\p p_t}{\p J_v}  } \\
 \displaystyle{\frac{\p p_t}{\p \Lz}  }
\end{pmatrix}
\endgroup
&=
\begingroup
\renewcommand*{\arraystretch}{2}
 \begin{pmatrix}
\Omega_u  & \Omega_v &\Omega_{\phi}   \\
 \displaystyle{\frac{\p I_3}{\p J_u}  }   &\displaystyle{\frac{\p I_3}{\p J_v}  } &\displaystyle{\frac{\p I_3}{\p \Lz}  }  \\
0& 0&1  
\end{pmatrix}^{\rT}
\endgroup
\begingroup
\renewcommand*{\arraystretch}{2}
 \begin{pmatrix}
 \displaystyle{\frac{\p p_t}{\p E}  }     \\
 \displaystyle{\frac{\p p_t}{\p I_3}  }    \\
 \displaystyle{\frac{\p p_t}{\p \Lz}  }   
\end{pmatrix}
\endgroup.
\end{align}

\section{Prolate spheroidal Basis elements}
\label{app:basis}

The Robijn basis elements for prolate spheroidal coordinates $(\xi,\eta,\phi)$ take the form  \citep[see section 3.2 of][]{Robijn1996b}
\begin{subequations}
\begin{align}
 \psi^{\ell m n}(\br) &=\frac{\sqrt{4\pi G }}{\Delta}F^{\ell m n}(\xi) \Phi^{\ell m}(\eta) \re^{\ri m \phi},  \\
\rho^{\ell m n}(\br) &=\frac{1} {\sqrt{4\pi G }\Delta}\frac{D^{\ell m n}(\xi) }{ \,c_{\Delta}(\xi,\eta)} \Phi^{\ell m}(\eta) \re^{\ri m \phi}, 
\end{align}
\end{subequations}
where $\ell=0,1,2,...$, $m=-\ell,...,\ell$, and $n=(0,)1,2,...$ (the 0-th order element appearing only when $m=0$).  Here, the density elements are related to the potential elements through the relation 
\begin{equation}
 \nabla^2_{\xi, \ell, m} F^{\ell m n}= D^{\ell m n}  ,
\end{equation}
where
\begin{equation}
\nabla^2_{\xi,\ell,m} = \frac{\rd}{\rd \xi} (\xi^2-1) \frac{\rd}{\rd \xi} - \ell(\ell+1) -\frac{m^2}{\xi^2-1}.
\end{equation}
We have also defined $c_{\Delta}(\xi,\eta)=\Delta^2(\xi^2-\eta^2)$.
 In particular, the $\eta$-components take the form $ \Phi^{\ell m}(\eta)=n^{\ell m} P_{\ell}^m(\eta)$, where $ P_{\ell}^m$ are the associated Legendre polynomials \citep[see, e.g.,][]{Abramowitz72} and 
\begin{equation}
n^{\ell m} = \sqrt{\frac{2\ell+1}{4\pi} \frac{(\ell-|m|)!}{(\ell+|m|)!}} (-1)^{\max(0,m)}.
\end{equation}
is the normalization constant. The normalization comes from that of the spherical harmonics $Y_{\ell}^m(v,\phi)$, which can be written as $Y_{\ell}^m(v,\phi)=\Phi^{\ell m}(\eta) \re^{\ri m \phi}$, with $\eta = \cos v$. They are normalized to
\begin{equation}
\int \rd v \rd \phi \sin v \, |Y_{\ell}^m(v,\phi)|^2 = 1.
\end{equation}
They may be called \textit{spheroidal harmonics} in this context. 
Now, we define the scalar product
\begin{equation}
\langle \psi^{(p)} ,  \psi^{(q)}  \rangle = -\frac{1}{4\pi G} \int \rd \br\, \psi^{(p)*}  \nabla^2 \psi^{(q)} .
\end{equation}
Requiring
the basis elements $(\psi^{(p)},\rho^{(p)})$ to be bi-orthogonal, i.e., satisfying the relation
\begin{equation}
\int \rd \br\, \psi^{(p)*}(\br) \rho^{(q)}(\br) = - \delta_{pq},
\end{equation}
means that $\{\psi^{(p)}\}_p$ should be orthonormal with respect to that scalar product.

Since the spheroidal harmonics are an orthonormal basis on the spheroid $(\eta,\phi)$, we only have to orthonormalize the radial $\xi$-elements for each fixed harmonics $(\ell,m)$. Thus, we should consider the radial inner product \begin{equation}
\langle F^{\ell m k}, F^{\ell m n} \rangle = -  \frac{1}{\Delta} \int_{1}^{\infty} \rd \xi \, F^{\ell m k}(\xi)^{*} D^{\ell m n}(\xi).
\end{equation}
Then, the full scalar product is related to the radial scalar product through
\begin{equation}
\langle \psi^{(p)} ,  \psi^{(q)}  \rangle = \delta_{\ell^p}^{\ell^q}\delta_{m^p}^{m^q} \langle F^{\ell^p m^p n^p}, F^{\ell^p m^p n^q} \rangle  .
\end{equation}

\subsection{Radial elements construction}

It is preferable to choose the form of the radial elements so that the inner products are easy to compute analytically. To that purpose, we use the form obtained by \citet{Robijn1996b}\footnote{Using their notations, we set  $p = h = 1$. }
\begin{equation}
\label{eq:Flmn}
F^{\ell m n}(\xi) = \frac{1}{\xi+1} \bigg( \frac{\xi-1}{\xi+1}\bigg)^n.
\end{equation}
Then, the associated density elements are given by
\begin{align}
 D^{\ell m n}(\xi) &\! =\! \frac{(\xi-1)^{n-1}}{(\xi+1)^{n+3}} \big( C_{0,n} \!+\! C_{1,n}(\xi-1) 
\!+\!C_{2,n}(\xi-1)^2 \big) \notag\\
&- \!\frac{(\xi-1)^{n-1}}{(\xi+1)^{n+1}} \bigg(\ell(\ell+1)(\xi-1) \! +\!  \frac{m^2}{\xi+1}\bigg), \hspace{-1mm}
\end{align}
where we defined the constants
\begin{subequations}
\begin{align}
C_{0,n} &= 8 n^2,\\
C_{1,n} &=2 \big( 1+2 n(n+1) -3 (2n+1)  \big) , \\
C_{2,n} &= 2  \big( 1 - 2(n+1) \big).
\end{align}
\end{subequations}
Note that the $n=0$ radial element should only be used if $m=0$.

\subsection{Computing the inner product}

The inner product can be written in the form 
\begin{align}
- \Delta \langle F^{\ell m k}, F^{\ell m n} \rangle 
&= C_{0,n}\,I_1(k+n-1,k+n+4) \notag \\
&+C_{1,n} \,I_1(k+n,k+n+4)     \notag\\
&+C_{2,n}\,I_1(k+n+1,k+n+4)     \notag\\
& -\ell(\ell+1)\,I_1(k+n,k+n+2)\notag\\
&-m^2 \,I_1(k+n-1,k+n+3) ,\notag
\end{align}
where 
\begin{align}
I_1(p_1,p_2) &= \int_{1}^{\infty} \rd \xi \frac{(\xi-1)^{p_1}}{(\xi+1)^{p_2}}. \end{align}
Following appendix~{B} of \citet{Robijn1996b}, we can compute these functions analytically. Indeed, 
\begin{align}
I_1(p_1,p_2)&= \mathrm{B}(p_1+1,p_2-p_1-1)\, 2^{p_1-p_2+1},
\end{align}
where $\mathrm{B}(x,y)$ is the Beta function.

\subsection{Radial basis orthogonalization}

Consider the non-orthogonal radial basis elements $\{\tF^i\}_i$. We define the Gram matrix $\boldsymbol{\mathcal{E}}^{\ell m}$ by letting
\begin{equation}
\langle \tF^{\ell m k},\tF^{\ell m n} \rangle  =  \mathcal{E}^{\ell m}_{kn}.
\end{equation}
Then, for a given harmonics $(\ell,m)$,  we can write the scalar product of any radial functions $G_1, G_2$ as 
\begin{equation}
\langle G_1,G_2 \rangle = G_1 ^T \boldsymbol{\mathcal{E}}^{\ell m} G_2.
\end{equation}
Let $\mathbf{R}$ be the Cholesky factorization of $ \boldsymbol{\mathcal{E}}^{\ell m} $. Then  $ \boldsymbol{\mathcal{E}}^{\ell m} =\mathbf{R}^T \mathbf{R}$, and we let $\mathbf{Y}=\mathbf{R}^{-1}$. It follows that
\begin{equation}
\mathbf{Y}^T \boldsymbol{\mathcal{E}}^{\ell m} \mathbf{Y} = \mathbf{Y}^T \mathbf{R}^T \mathbf{R} \mathbf{Y} = ( \mathbf{R} \mathbf{Y} )^T \mathbf{R} \mathbf{Y} = \bI.
\end{equation}
Therefore, the columns of $\mathbf{Y}$ form the (triangular) orthogonalization of the basis elements $\{\bar{F}^i\}_i$, and we have
\begin{equation}
\bar{F}^i = [Y^T \tF]_i= \sum_{j=n_0}^{i} Y_{j i} \tF^j ,
\end{equation}
where $n_0=0$ (resp.  $n_0=1$) if $m=0$ (resp. $m>0$).
We drop the bar notation from now on in order not to overload notations. As an example, the Gram--Schmidt coefficients for $(\ell,m)\!=\!(0,0)$ for $n\!=\!0,...,4$ are given by \begin{align}
\mathbf{Y} \!=\! \sqrt{\Delta}
\begingroup
\renewcommand*{\arraystretch}{1.2}
 \begin{pmatrix}
 2 & -\sqrt{2} & \frac{2}{\sqrt{3}} & -1 & \frac{2}{\sqrt{5}} \\
 0 & 3 \sqrt{2} & -\frac{16}{\sqrt{3}} & 15 & -\frac{48}{\sqrt{5}}\\
 0 & 0 & \frac{20}{\sqrt{3}} & -45 & \frac{252}{\sqrt{5}} \\
 0 & 0 & 0 & 35 & -\frac{448}{\sqrt{5}}  \\
 0 & 0 & 0 & 0 & \frac{252}{\sqrt{5}}
\end{pmatrix}.
\endgroup
\end{align}

In practice, all matrix manipulations are carried out using exact integer arithmetic and symbolic computation in \texttt{Mathematica}.\footnote{The GitHub repository, SPOCK, which contains this script, is publicly accessible at the following link: \href{https://github.com/KerwannTEP/SPOCK}{https://github.com/KerwannTEP/SPOCK}.} This allows us to obtain perfectly orthogonal basis elements with exact precision. Finally, we only need to compute the radial basis elements for  $m\geq0$ due to the $m \rightarrow -m$ symmetry.

\section{Linking  isochrone sphere to  Toomre disk}
\label{app:isochrone_to_toomre}

The Kuzmin--Kutuzov potential reduces to two special systems in some limits. On the one hand, in the spherical limit $a=c$, the potential reduces to the isochrone cluster \citep{Henon1960, Binney2008}, with its spherical potential given by
\begin{align}
\psi(r)=-\frac{G M}{a + \sqrt{a^2+r^2}}.
\end{align}
In particular, the distribution function reduces to the isotropic isochrone distribution \citepalias{Dejonghe1988}. On the other hand, in the flat limit $c=0$, the potential reduces to the Toomre disk \citep{Toomre1963}, described by the potential 
\begin{equation}
\label{eq:psi_flat}
\psi(R,z)=-\frac{G M}{\sqrt{R^2+(a+|z|)^2}},
\end{equation}
and the corresponding surface density
\begin{equation}
\label{eq:sigma_flat}
\Sigma(R)=\int_{-\infty}^{\infty}\hspace*{-3mm} \rd z\,\rho(R,z) = \frac{a M}{(2\pi)(R^2+a^2)^{3/2}}.
\end{equation}
One can show that equation~\eqref{eq:sigma_flat} is the flat limit of the Kuzmin--Kutuzov projected surface density. 

\begin{figure}
   \centering
\includegraphics[width=0.45 \textwidth]{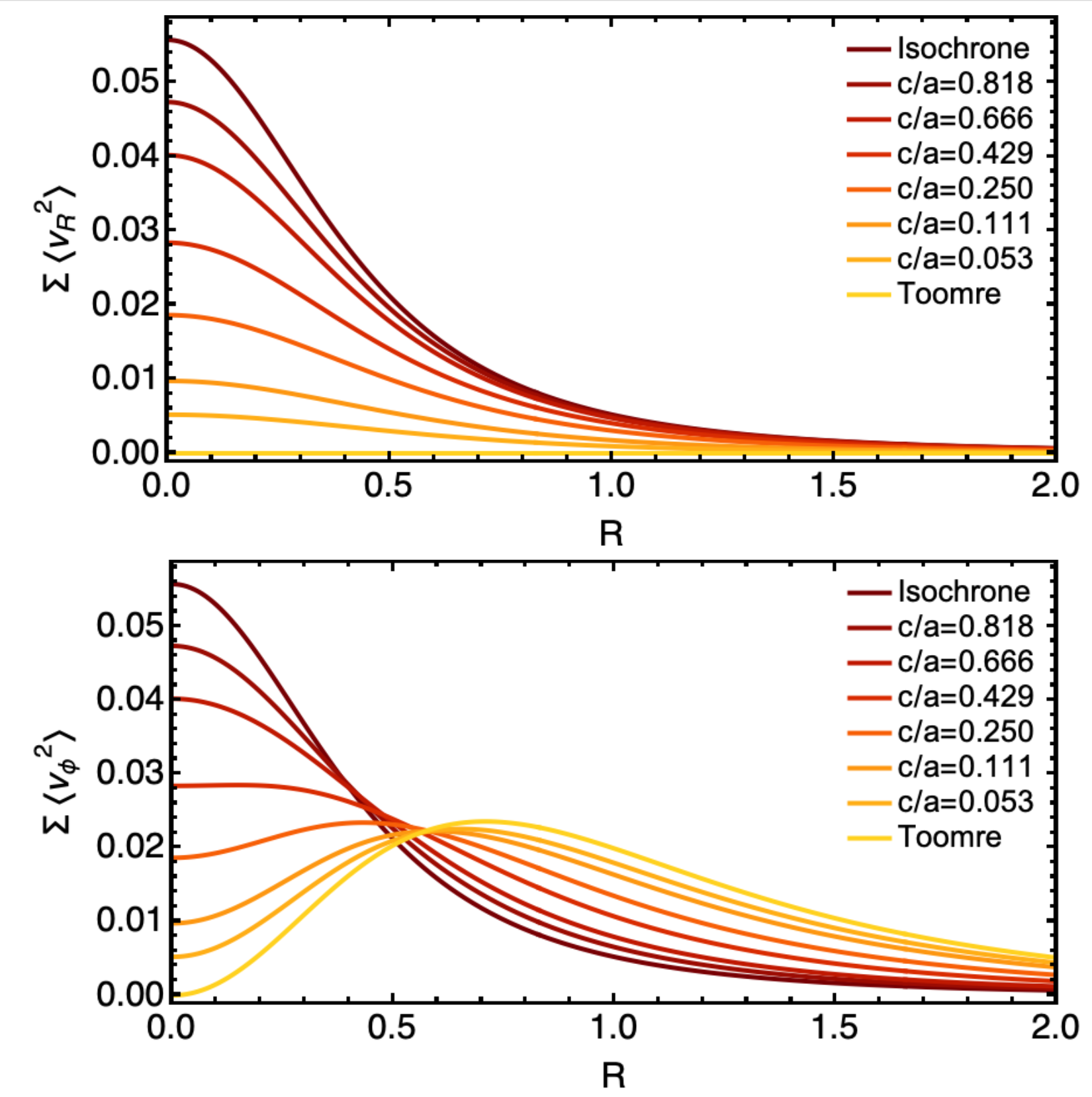}
   \caption{(Weighted) projected velocity dispersions (see, e.g., equations~{3.32} and {4.17} of  \citetalias{Dejonghe1988}) of a family of Kuzmin--Kutuzov potentials, as a function of radius $R$. \textit{Top panel:} $\Sigma \langle v_{R}^2 \rangle$; \textit{bottom panel:} $\Sigma \langle v_{\phi}^2 \rangle$. The flattened dispersions tend toward the ${m\!=\!\infty}$ Toomre disk dispersions \citep{Miyamoto1971}.
   }
   \label{fig:vphiSq}
\end{figure}

Since the Kuzmin--Kutuzov DF is a two-integral distribution, one always has $ {\langle v_R^2 \rangle \!=\!  \langle v_z^2 \rangle}$ \citepalias{Dejonghe1988}.  As such, in the flat limit, both these quantities go to zero together, so that the disc is radially cold.
In addition, we show in Fig.~\ref{fig:vphiSq} the transition of the projected angular velocity dispersion, $\langle v_{\phi}^2 \rangle$, between the spherical limit and the flat limit.
\begin{figure}
   \centering
   \includegraphics[width=0.45 \textwidth]{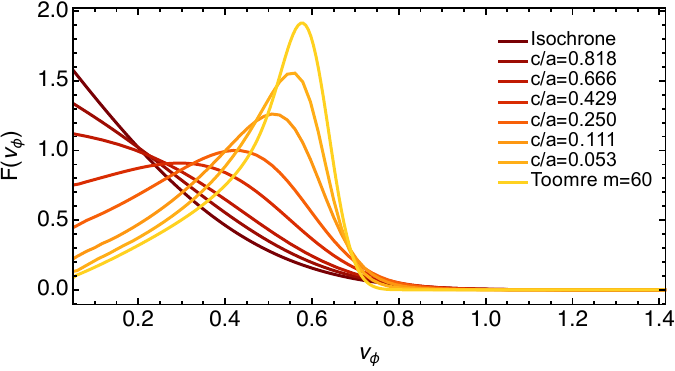}
   \vspace{-2mm}
   \caption{Distribution function of the angular velocity, $F(v_{\phi})$, as a function of flattening. The distribution function tends towards that of the ${m\!=\!\infty}$ Toomre disk \citep{Miyamoto1971}, which we approximate by the ${m\!=\!60}$ model.
   }
   \label{fig:DF_vphi}
\end{figure}
While the former stems from the isotropic isochrone, the latter can be obtained from the ${m\!=\!\infty}$ Miyamoto distribution \citep{Miyamoto1971}, which describes a Toomre disk containing only circular orbits. This is illustrated in Fig.~\ref{fig:DF_vphi}, where we show the behavior of the distribution function of the angular velocity, $F(v_{\phi})$, for a family of flattening parameters varying between the spherical limit and the flat limit.

\section{Computation of \texorpdfstring{$\psi_{\bk}^{(\lowercase{p})}(\bJ)$}{text}}
\label{app:response_matrix}

\subsection{From angle integration to elliptic integration}

We recall that  the (now) bi-orthogonal potential basis elements take the form
\begin{align}
\label{eq:basis_explicit}
\psi^{\ell m n}(\br) =\frac{ \sqrt{4\pi G}}{\Delta} F^{\ell m n}(\xi) Y_{\ell}^{m}(v,0) \,\re^{\ri m \phi}.
\end{align}
We let $\bk=(k_{u},k_{v},k_{\phi})$.
The Fourier transform of these elements is given by
\begin{equation}
\label{eq:FT_basic}
\psi_{\bk}^{(p)}(\bJ) = \int \frac{\rd \btheta}{(2\pi)^3} \psi^{(p)}(\br) \re^{-\ri \bk \cdot \btheta},
\end{equation}
where each angle is integrated over $[0,2\pi]$. Using equations~\eqref{eq:defangles}, we can switch between angle variables and $(u,v,\phi)$ variables.  Let us cut this integration into four parts
\begin{align}
\oint \rd \btheta = \underset{\substack{p_u>0 \\ p_v>0}}{\int} \rd \btheta 
+ \underset{\substack{p_u>0 \\ p_v<0}}{\int} \rd \btheta 
+ \underset{\substack{p_u<0 \\ p_v>0}}{\int} \rd \btheta 
+ \underset{\substack{p_u<0 \\ p_v<0}}{\int} \rd \btheta .
\end{align}
For each of these components, we can apply the bijective change of variables  $\btheta=(\theta_{u},\theta_{v},\theta_{\phi}) \mapsto (u,v,\phi)$, which yields
\begin{equation}
\underset{\substack{p_u \\ p_v}}{\int } \rd \btheta  [...] \!=\!  \underset{\hspace{-4mm} p_u}{{\int}_{\hspace{-1.8mm}u_0}^{u_1}} \hspace{-3mm}    \rd u  \underset{\hspace{-4mm} p_v}{{\int}_{\hspace{-1.8mm}v_0}^{v_1}}  \hspace{-3mm}    \rd v   \int_0^{2\pi}  \hspace{-3mm} \rd \phi  \,\bigg|\frac{\p(\theta_{u},\theta_{v},\theta_{\phi})}{\p(u,v,\phi)}\bigg|[...],
\end{equation}
where $p_u$ and $p_v$ have constant signs. Furthermore, using equations~\eqref{eq:pupv_derivatives}, the Jacobian takes the form
\begin{align}
\bigg |\frac{\p (\theta_u, \theta_v, \theta_{\phi})}{\p (u,v,\phi)} \bigg|
 &= \bigg| \frac{\p p_u}{\p J_u}\frac{\p p_v}{\p J_v}-   \frac{\p p_u}{\p J_v}  \frac{\p p_v}{\p J_u} \bigg|\\
&= \bigg | \big[J(\bw)^{-1}\big]^{\rT}\bigg | \, \bigg| \frac{\p p_u}{\p E}\frac{\p p_v}{\p I_3}-   \frac{\p p_u}{\p I_3}  \frac{\p p_v}{\p E} \bigg| \notag\\
&= \frac{\Delta^4}{\big |  J(\bw)\big |} \frac{\sinh^2 u + \sin^2 v}{|p_u\, p_v|},\notag
\end{align}
where
\begin{align}
\big |  J(\bw)\big | &= 
\bigg | \frac{\p J_u}{\p E}\frac{\p J_v}{\p I_3}- \frac{\p J_u}{\p I_3}\frac{\p J_v}{\p E}  \bigg|
.
\end{align}
It follows that the integrand of each $(u,v,\phi)$-integrals  is of the form
\begin{align}
[...] &= \Psi(u,v) \,\re^{\ri m \phi} \re^{-\ri \bk \cdot \btheta},\\
&= \Psi(u,v) \,\re^{\ri (m-k_{\phi}) \phi} \re^{-\ri ( \alpha_{\bk }[u] + \beta_{\bk }[v])},\notag
\end{align}
where $\Psi(u,v)$ is some function of $u$ and $v$ -- which depends neither on the sign of $p_u$ nor $p_v$ --  and we decomposed the angles into a $u$-part and a $v$-part using equations~\eqref{eq:defangles}. Indeed,  
\begin{align}
\label{eq:k_angles}
& k_{u} \theta_{u} +k_{v} \theta_{v}  +  k_{\phi}   (\theta_{\phi}-\phi)\\
=& \int_{u_0}^{u} \hspace{-2mm } \rd u' \bigg(k_{u} \frac{\p p_{u} }{\p J_{u}} + k_{v} \frac{\p p_{u} }{\p J_{v}} +  k_{\phi}   \frac{\p p_{u} }{\p \Lz} \bigg) \notag\\
+&\int_{v_0}^{v} \hspace{-2mm } \rd v'  \bigg(k_{u}  \frac{\p p_{v} }{\p J_{u}} +k_{v} \frac{\p p_{v} }{\p J_{v}}  +  k_{\phi}    \frac{\p p_{v} }{\p \Lz}\bigg)\notag \\
=& \alpha_{\bk }(u) + \beta_{\bk }(v) ,\notag
\end{align}
where we set $\bk=(k_u,k_v,k_{\phi})$.
We note that $ \alpha_{\bk }(u_1) =k_u \pi$ and  $ \beta_{\bk }(v_1) =k_v \pi$.
Now, carrying out the $\phi$ integration yields
\begin{equation}
\label{eq:FT_reduced}
\underset{\substack{p_u \\ p_v}}{\int } \rd \btheta  [...] \!=\! 2\pi \delta_{k_{\phi}}^{m}  \underset{\hspace{-4mm} p_u}{{\int}_{\hspace{-1.8mm}u_0}^{u_1}} \hspace{-3mm}    \rd u  \underset{\hspace{-4mm} p_v}{{\int}_{\hspace{-1.8mm}v_0}^{v_1}}  \hspace{-3mm}    \rd v     \,\Psi(u,v) \,\re^{-\ri ( \alpha_{\bk }  + \beta_{\bk } )}.
\end{equation}
Because regions with $p_u$ (resp. $p_v$) of opposing signs yield $\alpha_{\bk}$ (resp.  $\beta_{\bk}$) with opposing signs, we can pair the components two-by-two to obtain an explicitly real expression
\begin{align}
\oint \rd \btheta  [...]=  8\pi  \delta_{k_{\phi}}^{m}  {\int}_{ u_0}^{u_1} \hspace{-3mm}    \rd u   {\int}_{ v_0}^{v_1}  \hspace{-3mm}    \rd v     \,\Psi(u,v) \cos  \alpha_{\bk } \cos \beta_{\bk } .
\end{align}

\subsection{Fourier transform of the basis elements}

We let 
\begin{align}
\mathcal{J}(u,v) &=  \frac{\Delta^4(\sinh^2 u + \sin^2 v)}{\big |  J(\bw)\big |} \\
&=   \mathcal{J}_u(u)+ \mathcal{J}_v(v) . \notag
\end{align}
Then, combining equations~\eqref{eq:basis_explicit}, \eqref{eq:FT_basic} and \eqref{eq:FT_reduced} yields
\begin{align}
\psi_{\bk}^{\ell m n}(\bJ) 
&=\frac{ \sqrt{4\pi G}}{\Delta}\delta_{k_{\phi}}^{m} W_{\bk}^{\ell m n}(\bJ),
\end{align}
where we defined
\begin{align}
W_{\bk}^{\ell m n}(\bJ)
&= \frac{1 }{\pi^2}\int_{u_0}^{u_1} \hspace{-1mm}\frac{ \rd u}{p_u} \,F^{\ell m n}(\xi) \cos( \alpha_{\bk} )\\
&\times \int_{v_0}^{v_1} \hspace{-1mm} \frac{\rd v}{p_v}\, \mathcal{J} (u,v)\,Y_{\ell}^{m}(v,0) \,\cos(\beta_{\bk}  ).\notag
\end{align}

\subsection{Computation of  \texorpdfstring{$\alpha_{\bk}$}{text} and  \texorpdfstring{$\beta_{\bk}$}{text}}

Let use detail an efficient method to compute $\alpha_{\bk}$ and $\beta_{\bk}$.
Using equations~\eqref{eq:freq_matrix}, \eqref{eq:grad_mat}, \eqref{eq:relation_grads} and \eqref{eq:k_angles}, we have the relation
\begin{align}
\begingroup
\renewcommand*{\arraystretch}{2.0}
 \begin{pmatrix}
 \displaystyle{ \alpha_{\bk }(u)  }  \\
 \displaystyle{ \beta_{\bk }(v) }   \\
k_{\phi} \pi  
\end{pmatrix}^{\rT}
\endgroup
\!\!\!&= 
\!
\bk^{\rT}\!\big[J(\bw)^{-1}\big]^{\rT} \!  
\begingroup
\renewcommand*{\arraystretch}{2.0}
 \begin{pmatrix}
 \displaystyle{\int_{u_0}^u \hspace*{-2mm} \rd u' \,\frac{\p p_u}{\p E}  }  & \displaystyle{\int_{v_0}^v \hspace*{-2mm} \rd v' \,\frac{\p p_v}{\p E}  } & 0 \\
\displaystyle{\int_{u_0}^u \hspace*{-2mm} \rd u' \,\frac{\p p_u}{\p I_3}  }    & \displaystyle{\int_{v_0}^v \hspace*{-2mm} \rd v' \,\frac{\p p_v}{\p I_3}  }  & 0\\
\displaystyle{\int_{u_0}^u\hspace*{-2mm}  \rd u' \,\frac{\p p_u}{\p \Lz}  }  & \displaystyle{\int_{v_0}^v \hspace*{-2mm} \rd v' \,\frac{\p p_v}{\p \Lz}  } & \pi 
\end{pmatrix}
\endgroup
, \notag
\end{align} 
where $J(\bw)$ is the Jacobian from equation~\eqref{eq:Jacobian}. For $u=u_1$ and $v=v_1$, this reduces to
\begin{align}
\begingroup
\renewcommand*{\arraystretch}{2.0}
 \begin{pmatrix}
 \displaystyle{ \alpha_{\bk }(u_1)  }  \\
 \displaystyle{ \beta_{\bk }(v_1) }   \\
k_{\phi} \pi  
\end{pmatrix}^{\rT}
\endgroup
 &= 
\pi \, \bk^{\rT} \big[J(\bw)^{-1}\big]^{\rT}  J(\bw)^{\rT}
 = 
\pi \, \bk^{\rT} 
.
\end{align}
Using the effective anomalies $\tilde{u}$ and $\tilde{v}$, we can rewrite the last matrix as
\begin{align}
\begingroup
\renewcommand*{\arraystretch}{2.0}
 \begin{pmatrix}
 \displaystyle{\int_{-\pi/2}^{\tilde{u}} \hspace*{-4mm} \rd \tilde{u}' \,\frac{\Delta^2 \sinh^2 u'}{\widetilde{p}_u}  }  & \displaystyle{\int_{-\pi/2}^{\tilde{v}} \hspace*{-4mm} \rd \tilde{v}' \,\frac{\Delta^2 \sin^2 v'}{\widetilde{p}_v}  } & 0 \\
-\displaystyle{\int_{-\pi/2}^{\tilde{u}} \hspace*{-4mm} \rd \tilde{u}' \,\frac{\Delta^2}{\widetilde{p}_u}  }    & \displaystyle{\int_{-\pi/2}^{\tilde{v}} \hspace*{-4mm} \rd \tilde{v}' \,\frac{\Delta^2}{\widetilde{p}_v}  }  & 0\\
-\displaystyle{\int_{-\pi/2}^{\tilde{u}}\hspace*{-4mm}  \rd \tilde{u}' \,\frac{\Lz}{\widetilde{p}_u \sinh^2 u'}  }  & -\displaystyle{\int_{-\pi/2}^{\tilde{v}} \hspace*{-4mm} \rd \tilde{v}' \,\frac{\Lz}{\widetilde{p}_v \sin^2 v'}  } & \pi 
\end{pmatrix}
\endgroup
.
\end{align}
Using backward integration, we let  $\alpha_{\bk}=k_u \pi - \alpha^{\rr}_{\bk}$ and $\beta_{\bk}=k_v \pi - \beta^{\rr}_{\bk}$. These new quantities read
\begin{align}
\begingroup
\renewcommand*{\arraystretch}{2.0}
 \begin{pmatrix}
 \displaystyle{ \alpha^{\rr}_{\bk }(u)  }  \\
 \displaystyle{ \beta^{\rr}_{\bk }(v) }   \\
0 
\end{pmatrix}^{\rT}
\endgroup
\!\!\!\!\!&= 
\!
\bk^{\rT}\!\big[J(\bw)^{-1}\big]^{\rT} \!  
\begingroup
\renewcommand*{\arraystretch}{2.0}
 \begin{pmatrix}
 \displaystyle{\int_{u}^{u_1} \hspace*{-2mm} \rd u' \,\frac{\p p_u}{\p E}  }  & \displaystyle{\int_{v}^{v_1} \hspace*{-2mm} \rd v' \,\frac{\p p_v}{\p E}  } & 0 \\
\displaystyle{\int_{u}^{u_1} \hspace*{-2mm} \rd u' \,\frac{\p p_u}{\p I_3}  }    & \displaystyle{\int_{v}^{v_1} \hspace*{-2mm} \rd v' \,\frac{\p p_v}{\p I_3}  }  & 0\\
\displaystyle{\int_{u}^{u_1}\hspace*{-2mm}  \rd u' \,\frac{\p p_u}{\p \Lz}  }  & \displaystyle{\int_{v}^{v_1} \hspace*{-2mm} \rd v' \,\frac{\p p_v}{\p \Lz}  } &0
\end{pmatrix}
\endgroup
. \notag
\end{align}
Given $\tilde{u}$ and $\tilde{v}$, we can rewrite the last matrix as
\begin{align}
\begingroup
\renewcommand*{\arraystretch}{2.0}
 \begin{pmatrix}
 \displaystyle{\int_{\tilde{u}}^{\pi/2} \hspace*{-4mm} \rd \tilde{u}' \,\frac{\Delta^2 \sinh^2 u'}{\widetilde{p}_u}  }  & \displaystyle{\int_{\tilde{v}}^{\pi/2} \hspace*{-4mm} \rd \tilde{v}' \,\frac{\Delta^2 \sin^2 v'}{\widetilde{p}_v}  } & 0 \\
-\displaystyle{\int_{\tilde{u}}^{\pi/2} \hspace*{-4mm} \rd \tilde{u}' \,\frac{\Delta^2}{\widetilde{p}_u}  }    & \displaystyle{\int_{\tilde{v}}^{\pi/2} \hspace*{-4mm} \rd \tilde{v}' \,\frac{\Delta^2}{\widetilde{p}_v}  }  & 0\\
-\displaystyle{\int_{\tilde{u}}^{\pi/2}\hspace*{-4mm}  \rd \tilde{u}' \,\frac{\Lz}{\widetilde{p}_u \sinh^2 u'}  }  & -\displaystyle{\int_{\tilde{v}}^{\pi/2} \hspace*{-4mm} \rd \tilde{v}' \,\frac{\Lz}{\widetilde{p}_v \sin^2 v'}  } & 0
\end{pmatrix}
\endgroup
.
\end{align}

\subsection{Computation of \texorpdfstring{$W_{\bk}^{(\lowercase{p})}(\bJ)$}{text}}

Using the separability of $\mathcal{J}$, we have
\begin{align}
W_{\bk}^{\ell m n} \!
&= \! \int_{u_0}^{u_1} \hspace{-1.5mm}\frac{ \rd u}{\pi} \,\frac{\mathcal{J}_u F^{\ell m n}  \cos( \alpha_{\bk}  ) }{p_u}\hspace*{-1.5mm}
  \int_{v_0}^{v_1} \hspace{-1mm} \frac{\rd v}{\pi}\,\frac{   \,Y_{\ell}^{m}(v,0) \,\cos(\beta_{\bk}  )}{p_v}\notag \\
&+ \!\! \int_{u_0}^{u_1} \hspace{-1.mm}\frac{ \rd u}{\pi} \,\frac{F^{\ell m n} \cos( \alpha_{\bk}  ) }{p_u}\hspace*{-1.mm}
  \int_{v_0}^{v_1} \hspace{-1mm} \frac{\rd v}{\pi}\,\frac{ \mathcal{J}_v  \,Y_{\ell}^{m}(v,0) \,\cos(\beta_{\bk}  )}{p_v},
  \notag
\end{align}
which involves four separate one-dimensional integrals. Using effectives anomalies $\tilde{u}$ and  $\tilde{v}$, they can be expressed using non-singular integrands as
\begin{align}
W_{\bk}^{\ell m n} \!
&\!= \! \!\int_{-\frac{\pi}{2}}^{\frac{\pi}{2}} \hspace{-1mm}\frac{ \rd \tilde{u}}{\pi} \,\frac{\mathcal{J}_u F^{\ell m n} \! \cos( \alpha_{\bk}  ) }{\tilde{p}_u}\hspace*{-1mm}
 \int_{-\frac{\pi}{2}}^{\frac{\pi}{2}}\hspace{-1mm} \frac{\rd \tilde{v}}{\pi}\,\frac{   \,Y_{\ell}^{m}(v,0) \!\cos(\beta_{\bk}  )}{\tilde{p}_v}\notag \\
&\!+\! \! \int_{-\frac{\pi}{2}}^{\frac{\pi}{2}} \hspace{-1mm}\frac{ \rd \tilde{u}}{\pi} \,\frac{F^{\ell m n} \!\cos( \alpha_{\bk}  ) }{\tilde{p}_u}\hspace*{-1mm}
 \int_{-\frac{\pi}{2}}^{\frac{\pi}{2}} \hspace{-1mm} \frac{\rd \tilde{v}}{\pi}\,\frac{ \mathcal{J}_v  \,Y_{\ell}^{m}(v,0) \!\cos(\beta_{\bk}  )}{\tilde{p}_v}, \notag
\end{align}
where the angles $\alpha_{\bk} $ and $\beta_{\bk} $ can be computed on the fly following the previous section's scheme. In practice, we compute these integrals using a backward scheme.
\subsection{Convergence study}
\begin{figure}
    \centering
   \includegraphics[width=0.45 \textwidth]{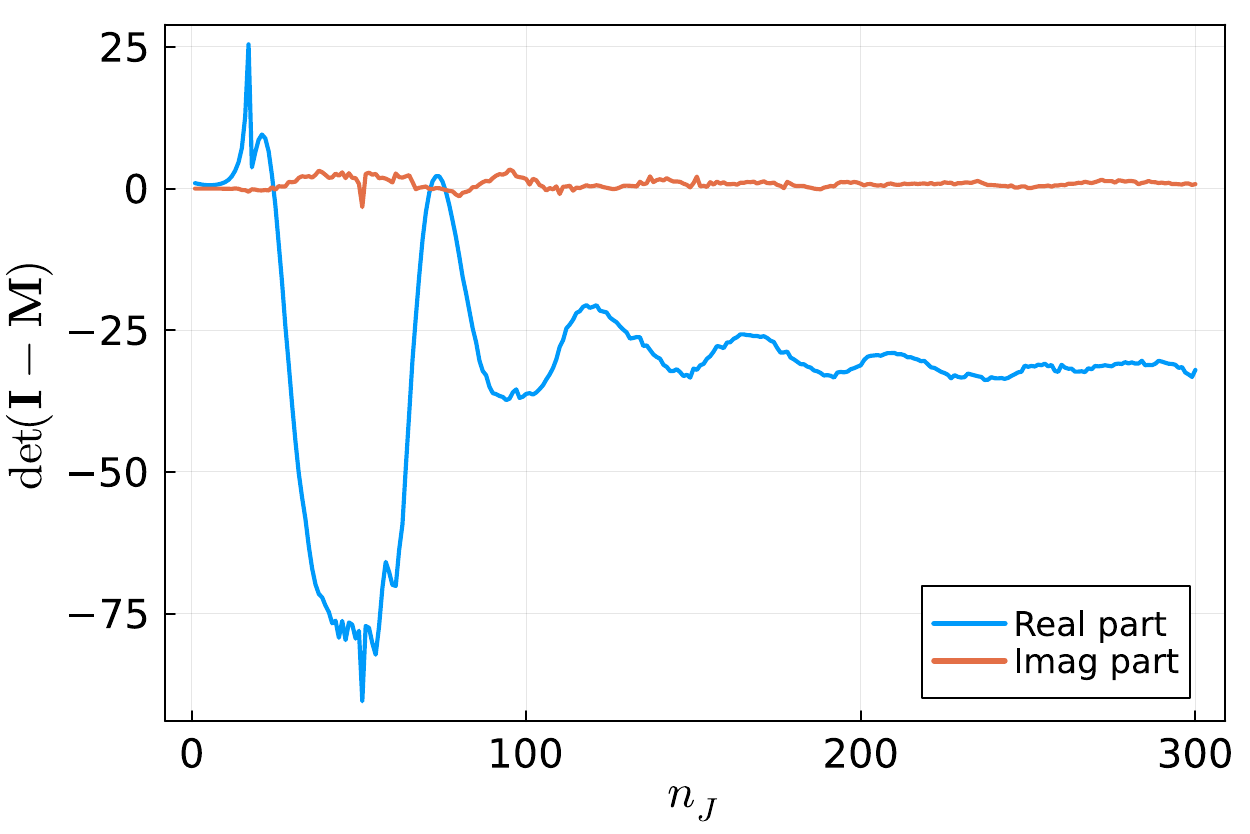}
   \caption{Convergence of $\det(\bI-\bM[\omega])$, for $m=2$, w.r.t. the sampling numbers, $n_J$, of the $u$ and $v$ action variables. The plotted value is taken to be the average between a sampling number of the $\Lz$ action of $80$  and of $81$ -- to increase convergence speed -  while we set $a=0.85$ and $\omega = 0.01 \ri$.
   }
   \label{fig:CV_detN_nbJ}
\end{figure}

Figure~\ref{fig:CV_detN_nbJ} presents the convergence of $\det(\bI-\bM[\omega])$ w.r.t. the sampling numbers, $n_J$, of the $u$ and $v$ action variables in equation~\eqref{eq:Mpq_m}. The action space integral itself is performed by transforming the infinite domain $[0,\infty)\!\times\![0,\infty)\!\times\!(-\infty,\infty)$ into the  compact domain $[0,\pi/2)\!\times\![0,\pi/2)\!\times\!(-\pi/2,\pi/2)$ by using the change of variables
\begin{align}
  J_i \mapsto \tilde{J}_i = \tan^{-1}\bigg(\frac{J_i}{\sqrt{G M (a+c)}}\bigg),
  \end{align}
for each action variables. The integration itself is then carried out using mid-point sampling. 
From this figure we assume convergence for $n_J=256$, which we take for our computation. As for $\Lz$, we use in this paper $n_{\Lz}=128$.

\section{Introducing rotation}
\label{app:rotation}

\subsection{Distribution function:  Lynden-Bell daemon}

Using the Lynden-Bell parametrization \citep{LyndenBell1960}, the density profile of the cluster does not depend on the rotation parameter $\alphar$. Its derivatives w.r.t. the action variables read
\begin{subequations}
\begin{align}
\frac{\p \Frot}{\p J_u} &\!=\! \frac{\p F}{\p J_u} (1\! +\! \alphar\, \sgn [\Lz]),\\
\frac{\p \Frot}{\p J_v} &\!=\! \frac{\p F}{\p J_v} (1 \!+\! \alphar\, \sgn [\Lz]),\\
\frac{\p \Frot}{\p \Lz} &\!=\! \frac{\p F}{\p \Lz} (1 \!+\! \alphar\, \sgn [\Lz])\! +\!2 \alphar  F(\bJ) \deltaD (\Lz)\,,
\end{align}
\end{subequations}
where $F(\bJ)=F(E[J_u,J_v,\Lz],\Lz)$. Therefore, the computation of the response matrix requires slight modifications. 

Most notably, one must take into account the Dirac delta contribution, whose implementation should be performed carefully. Indeed, the third frequency $\Omega_z$ has two distinct left-sided and right-sided limits at $\Lz \rightarrow 0$. We can remedy this issue by recalling that integration over $\deltaD$ should be understood as the limit 
\begin{align}
 \int_{-\infty}^{\infty}\hspace*{-3mm}  \rd x \, g(x) \deltaD(x) = \lim_{a \rightarrow 0^{+}}  \int_{-\infty}^{\infty}\hspace*{-3mm}  \rd x \, g(x) \eta_a(x).
\end{align}
over a set of functions $g(x)$.
Here, ${\eta_a\!  \rightarrow \!\deltaD}$ converges in the distributional sense, with ${\eta_a(x)\!=\!\eta(x/a)/a}$ where $\eta$ is an even function integrating to 1. Then 
\begin{align}
&\int_{-\infty}^{\infty}\hspace*{-3mm}  \rd x \, g(x) \eta_a(x)= \int_{-\infty}^{\infty}\hspace*{-3mm}  \rd y \, g(a y) \eta(y)\\
&=\int_{-\infty}^{0}\hspace*{-3mm}  \rd y \, g(a y) \eta(y)+\int_{0}^{\infty}\hspace*{-3mm}  \rd y \, g(a y) \eta(y) \notag\\
& \rightarrow \int_{-\infty}^{0}\hspace*{-3mm}  \rd y \, g(0^{-}) \eta(y)+\int_{0}^{\infty}\hspace*{-3mm}  \rd y \, g(0^{+}) \eta(y) .\notag
\end{align}
Since $\eta$ is even and integrates to 1, it follows that 
\begin{align}
\label{eq:dirac_lim}
\int_{-\infty}^{\infty}\hspace*{-3mm}  \rd x \, g(x) \deltaD(x)  = \frac{g(0^{+})+g(0^{-})}{2}.
\end{align}

\subsection{Response matrix}

The rotational response matrix elements can be decomposed into two contributions
\begin{equation}
\rM_{pq}^m(\omega) = \rM_{pq}^{m,a}[\omega] +\alphar \big( \rM_{pq}^{m,b}[\omega]+ \rM_{pq}^{m,c}[\omega] \big),
\end{equation}
where \begin{subequations}
\begin{align}
\rM_{pq}^{m,a}(\omega)&\!=\! \frac{ 32\pi^4 G }{\Delta^2} \hspace*{-5mm} \sum_{\substack{k_u,k_v \\ \ell^p+m+k_v\, \mathrm{even}\\ \ell^q+m+k_v\, \mathrm{even}}} \hspace*{-4mm} \!\!\int \!\rd \bJ \frac{\bk \!\cdot\! \p F/\p \bJ}{\omega - \bk \cdot \bOmega }    W_{\bk}^{(p) }   W_{\bk}^{(q) } ,\notag\\
\rM_{pq}^{m,b}(\omega)&\!=\! \frac{ 32\pi^4 G }{\Delta^2} \hspace*{-5mm} \sum_{\substack{k_u,k_v \\ \ell^p+m+k_v\, \mathrm{even}\\ \ell^q+m+k_v\, \mathrm{even}}} \hspace*{-4mm} \!\!\int \!\rd \bJ \frac{\bk \!\cdot\! \p F/\p \bJ}{\omega - \bk \cdot \bOmega }  \sgn[\Lz]  W_{\bk}^{(p) }   W_{\bk}^{(q) } ,\notag\\
\rM_{pq}^{m,c}(\omega)&\!=\!  \frac{ 64  \pi^4 G }{\Delta^2} \hspace*{-5mm} \sum_{\substack{k_u,k_v \\ \ell^p+m+k_v\, \mathrm{even}\\ \ell^q+m+k_v\, \mathrm{even}}} \hspace*{-4mm} \!\!\int \!\rd \bJ \frac{m F(\bJ)}{\omega - \bk \cdot \bOmega } \deltaD(\Lz)  W_{\bk}^{(p) }   W_{\bk}^{(q) } .\notag
\end{align}
\end{subequations}
Using the Dirac delta function, we can reduce the 3D integral to a 2D integral by applying equation~\eqref{eq:dirac_lim}
\begin{align}
\rM_{pq}^{m,c}(\omega)&\!=\!  \frac{ 64  \pi^4 G }{\Delta^2} \hspace*{-5mm} \sum_{\substack{k_u,k_v \\ \ell^p+m+k_v\, \mathrm{even}\\ \ell^q+m+k_v\, \mathrm{even}}} \hspace*{-4mm} \!\!\int \!\rd J_u \rd J_v  \bigg[ \frac{m F(\bJ)}{\omega - \bk \cdot \bOmega } W_{\bk}^{(p) }   W_{\bk}^{(q) } \bigg]_{\rs}, \notag
\end{align}
where we defined the symmetric central value
\begin{equation}
\big[ g(\bJ) \big]_{\rs} = \lim_{\delta \Lz \rightarrow 0^+}\frac{g(J_u,J_v,\delta \Lz) + g(J_u,J_v,-\delta \Lz)}{2} . \notag
\end{equation}

\section{$z$-slice of the bending  mode's shape}
\label{app:comp_fig_bend_mode_shape}
\begin{figure}
    \centering
\includegraphics[width=0.4 \textwidth]{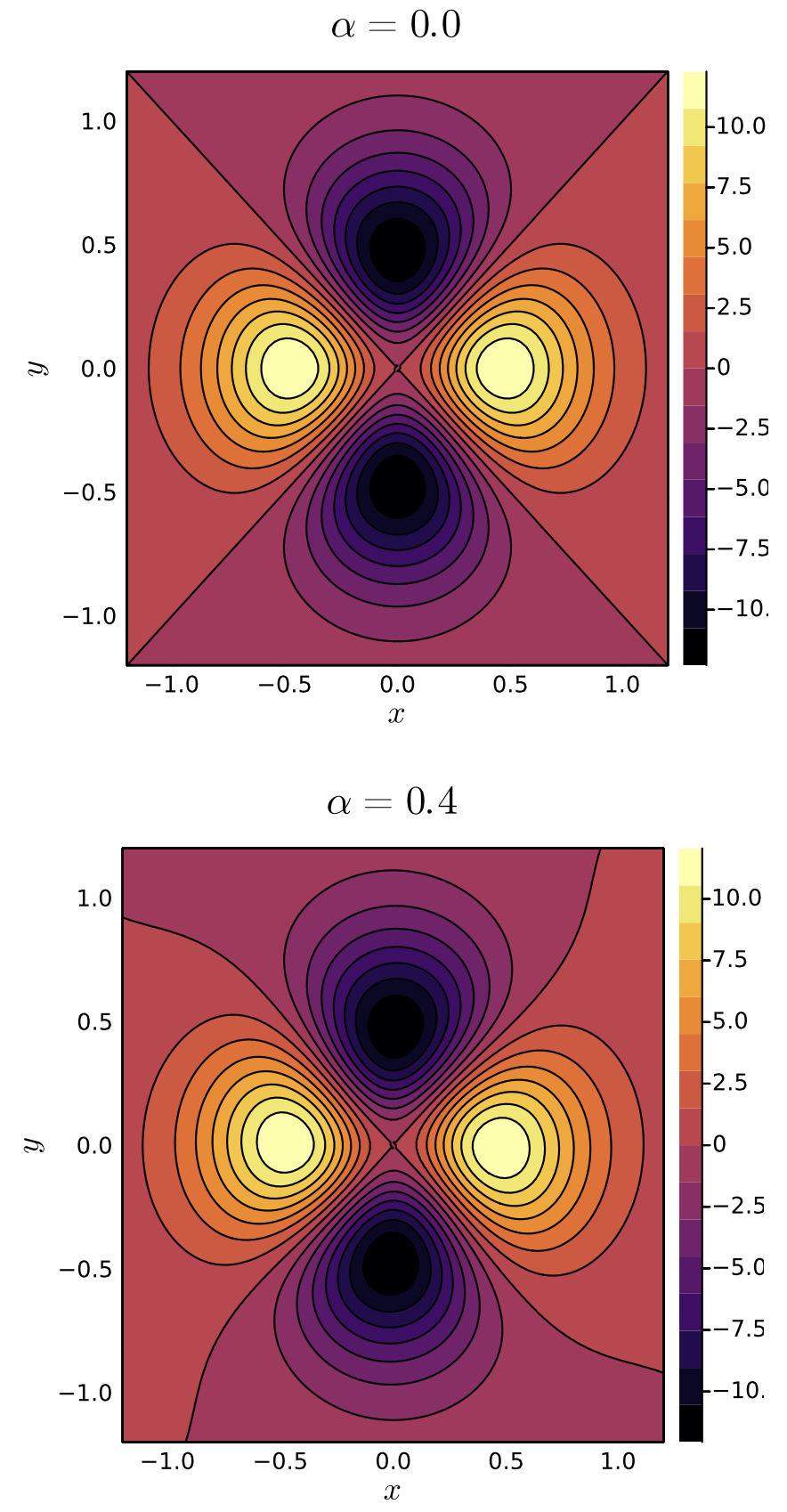}
\vspace{-1mm}
   \caption{Slice $z=0.1$ of the $m=2$ density modes of a $a=0.9$ cluster, for the non-rotating bending mode ($\alpha=0$, top panel, see Fig.~\ref{fig:mode_shape_m_2}), and a rotating bending mode ($\alpha=0.4$, bottom panel, see Fig.~\ref{fig:mode_shape_rot_m_2_slow}). Rotation only weakly impacts the shape of the mode.    
      }
   \label{fig:shape_bending_rot_slice}
\end{figure}

As a complement to Fig.~\ref{fig:mode_shape_rot_m_2_slow},
let us highlight  the impact of  rotation of the shape of bending mode by representing a $z$-slice of the mode in Fig.~\ref{fig:shape_bending_rot_slice}.
Rotation appears to have a small, but real impact on the mode's shape.

\section{Analytic continuation}
\label{app:analytic_continuation}

Let us consider a function $g(z)$, for which we have access to a set of $2N+1$ values $\{g_k\}_k=\{g(z_k)\}_k$ over a sampling $\{z_k\}_k$. Let us consider the approximation 
\begin{align}
g(z) = \frac{P(z)}{Q(z)} = \frac{\sum_{i=0}^{N} a_i z^i }{1+\sum_{i=1}^{N} b_i z^i }.
\end{align}
Although there exist recursive relations to compute the coefficients of $P$ and $Q$ \citep[see, e.g.,][]{StoerBulirsch1980}, one can directly compute those using linear algebra. Indeed, let us define $\brc=(a_0,...,a_N,b_1,...,b_N)$, $\brg=(g_1,...,g_{2N+1})$ and the matrix $\brA$ such that
\begin{align}
A_{kj}
&=
\begin{cases}
\hspace*{7mm}z_k^{j-1}\hspace*{4mm}, & {\mathrm{if}} \ j=1,...,N+1 ,\\
-g_k\, z_k^{j-N-1},&{\mathrm{otherwise}} .
\end{cases}
\end{align}
Then, one can recover the coefficients $\brc$ by solving the equation
\begin{align}
\brA \, \brc =  \brg.
\end{align}

\section{Maclaurin Spheroid}
\label{app:Maclaurin}

Self-gravitating fluids with uniform rotation are known to possess flattened equilibrium states: the so-called Maclaurin spheroids \citep{Chandrasekhar1969}.
Their  angular momentum, $ L$, and energy, $E$,   read as a function of eccentricity
\begin{subequations}
\begin{align}
 \hat{L} &=\! \frac{\sqrt{6}}{5} \frac{(\cos \gamma)^{-\tfrac{1}{6}}}{\sin \gamma} \sqrt{(1+2\cos^2 \gamma)\frac{\gamma}{\sin \gamma} \!-\! 3 \cos \gamma},\!\!\\
 \hat{E}&=\!-\frac{3}{10} \frac{(\cos \gamma)^{\tfrac{1}{3}}}{\sin^2 \gamma}  \bigg[(1-4 \cos^2 \gamma)\frac{\gamma}{\sin \gamma} \!+\! 3 \cos \gamma\bigg],\!\!
\end{align}
\end{subequations}
where $e=\sin \gamma=\sqrt{1-c^2/a^2}$, $\bar a= (a^2 c)^{1/3}$, $\hat{L}=L/\sqrt{G M^3 \bar{a}}$ and $\hat{E}=-a E/(G M^2)$. Therefore, it follows that its spin parameter is given by
\begin{equation}
\lambda_{\rr} =\hat{L} \sqrt{\hat{E}} \sqrt{\bar a/a } .
\end{equation}

\section{Impact of the discontinuity of the distribution function}
\label{app:LBD_disc}

\begin{figure}
   \centering
    \includegraphics[width=0.45 \textwidth]{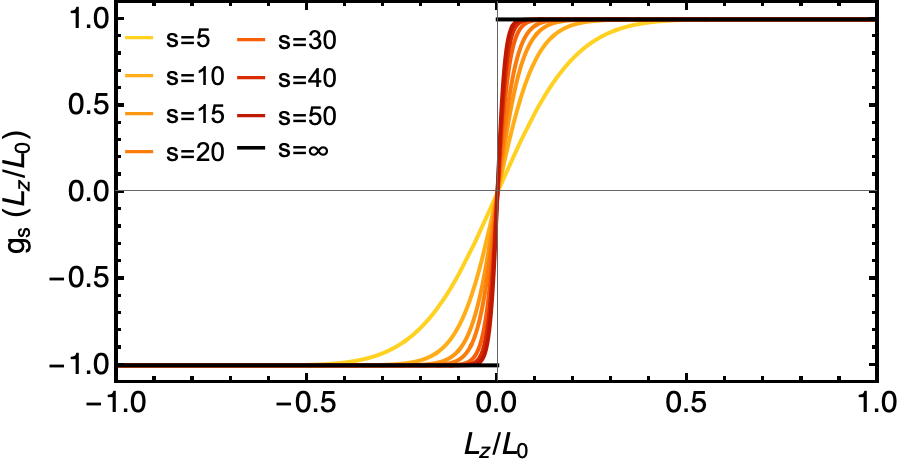}
   \caption{Family of functions $g_s(\Lz/L_0)$, defined by equation~\eqref{eq:def_gs}, for various values of $s$. These functions approach the sign function in the $s \rightarrow \infty$ limit, while keeping a smooth behavior, which allows us to study the impact of discontinuity in the location of the modes.
   }
   \label{fig:sgn_approx}
\end{figure}

\begin{figure}
   \centering
    \includegraphics[width=0.45 \textwidth]{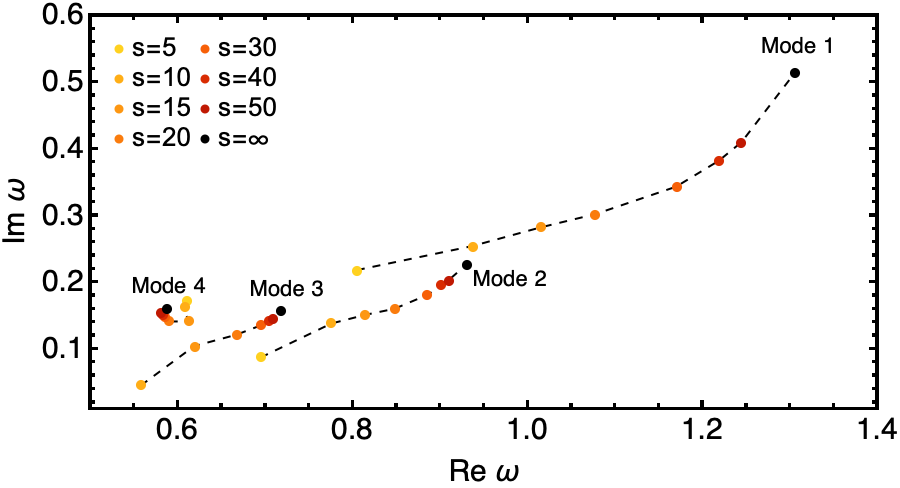}
   \caption{Impact of the smoothing parameter $s$ (see equation~\ref{eq:def_gs}) on the location of the $m=2$ bar modes for a $c/a=0.11$ cluster with maximal rotation $\alphar=1$ (see Fig.~\ref{fig:modes_subdominant}). For illustration purposes, we consider a truncated response matrix with parameters  $\ell_{\max}=20$, $n_{\max}=10$ and $k_{\max}=10$. As we consider distribution functions closer to the discontinuous LBD limit ($s \rightarrow \infty$), we can observe the modes of the smoothed DF converging towards the modes of the discontinuous DF.}
   \label{fig:sgn_approx_modes}
\end{figure}

To the probe the impact of the discontinuity of the $\sgn$ function in the Lynden-Bell parametrization (equation~\ref{eq:KKLB}), we consider a smooth approximation of the $\sgn$ function
\begin{align}
\label{eq:def_gs}
g_s(x)
&=
\begin{cases}
\hspace*{12.3mm}-1&{\mathrm{if}} \ x \leq -1,\\
\displaystyle{\tanh\bigg(\frac{s x}{1-x^2}\bigg)} & {\mathrm{if}} \ |x|<1 ,\\
 \hspace*{15mm}1 & {\mathrm{if}} \ x\geq 1 .
\end{cases}
\end{align}
such that $g_s(x) \rightarrow \sgn(x)$ as $s \rightarrow \infty$ (Fig.~\ref{fig:sgn_approx}).
It follows that
\begin{subequations}
\begin{align}
\frac{\p \Frot}{\p J_u} &\!=\! \frac{\p F}{\p J_u} \bigg(1\! +\! \alphar\, g_s \bigg[\frac{\Lz}{L_0}\bigg]\bigg),\\
\frac{\p \Frot}{\p J_v} &\!=\! \frac{\p F}{\p J_v}  \bigg(1\! +\! \alphar\, g_s \bigg[\frac{\Lz}{L_0}\bigg]\bigg),\\
\frac{\p \Frot}{\p \Lz} &\!=\! \frac{\p F}{\p \Lz}  \bigg(1\! +\! \alphar\, g_s \bigg[\frac{\Lz}{L_0}\bigg]\bigg)\! +\! \frac{\alphar  F(\bJ)}{L_0}g_s' \bigg[\frac{\Lz}{L_0}\bigg]\,,
\end{align}
\end{subequations}
where  $g_s'$  vanishes for $|x|\geq 1$, and reads
\begin{align}
g_s'(x)&= \bigg[\frac{2 s x^2}{(1-x^2)^2}+\frac{s}{1-x^2}\bigg]\,\mathrm{sech}^2\bigg(\frac{s x}{1-x^2}\bigg),
\end{align}
when $|x|<1$. Using a truncated response matrix for illustration purposes -- with parameters $\ell_{\max}=20$, $n_{\max}=10$ and $k_{\max}=10$ -- we show in Fig.~\ref{fig:sgn_approx_modes} the location of the $\alpha=1$ bar modes of the $c/a=0.11$ cluster, as a function of the smoothing parameter $s$. 
The locations of all the modes from the smooth DFs appear to converge towards those of the discontinuous LBD distribution function. As such, none of these modes appear to be purely the result of the discontinuity of the DF, proving that none of them are in fact edge modes.

In addition, Fig.~\ref{fig:sgn_approx_modes} clearly shows how much smoothing can impact the growth rates and pattern speeds of the modes. In particular, the location of the fastest mode (Mode 1) gets quite close to that of the discontinuous Mode 2 for the smoothing parameters $s \leq 10$. This prediction is consistent with the measurements made by  \citetalias{Sellwood1997}, who considered a smoothed DF whose behavior was qualitatively similar to our own.

\section{Subdominant  bending  and  bar  modes}
\label{app:bend_v_bar}

\begin{figure*}
   \centering
    \includegraphics[width=0.9 \textwidth]{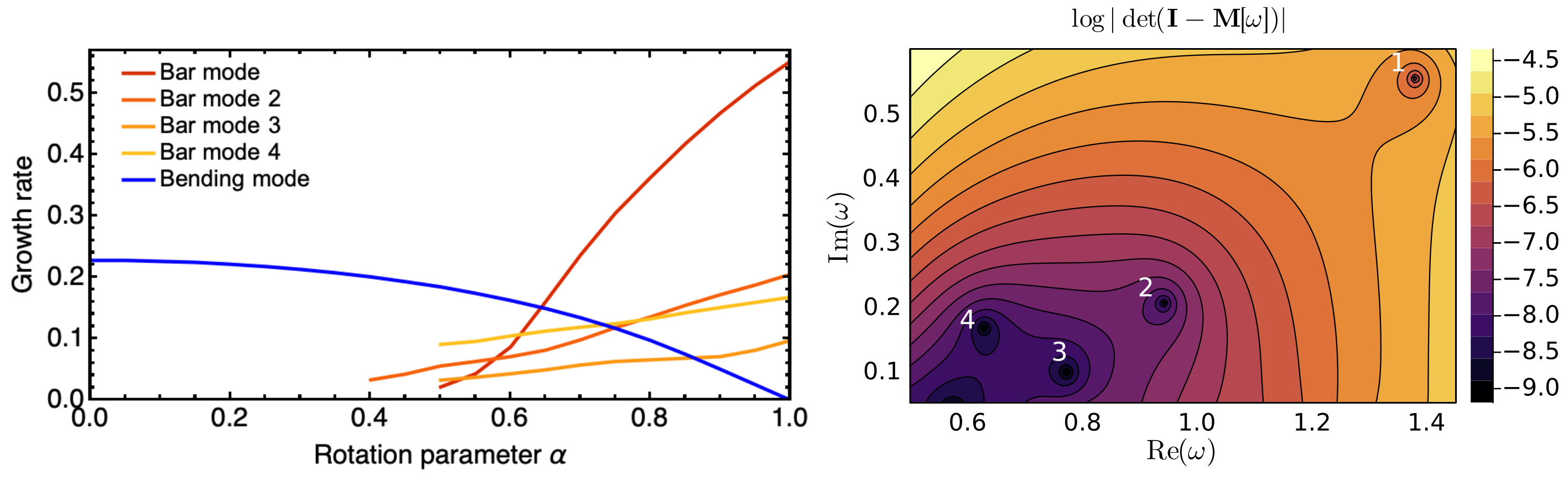}
\caption{\textit{Left panel: }Growth rate of the dominant bending modes (in blue) and  bar-growing mode (in red), as a function of rotation, for a clusters with flattening ratio $c/a\!=\!0.11$. We used the parameters $\ell_{\max}\!=\!30$ and $n_{\max}\!=\!20$. The bending mode is dominant for slowly rotating clusters, with a growth rate which goes to 0 for the maximally rotating cluster. \textit{Right panel:} Location in frequency space of the four bar modes of the maximally rotating cluster. }
   \label{fig:modes_subdominant}
\end{figure*}

\begin{figure*}
   \centering
\includegraphics[width=0.8 \textwidth]{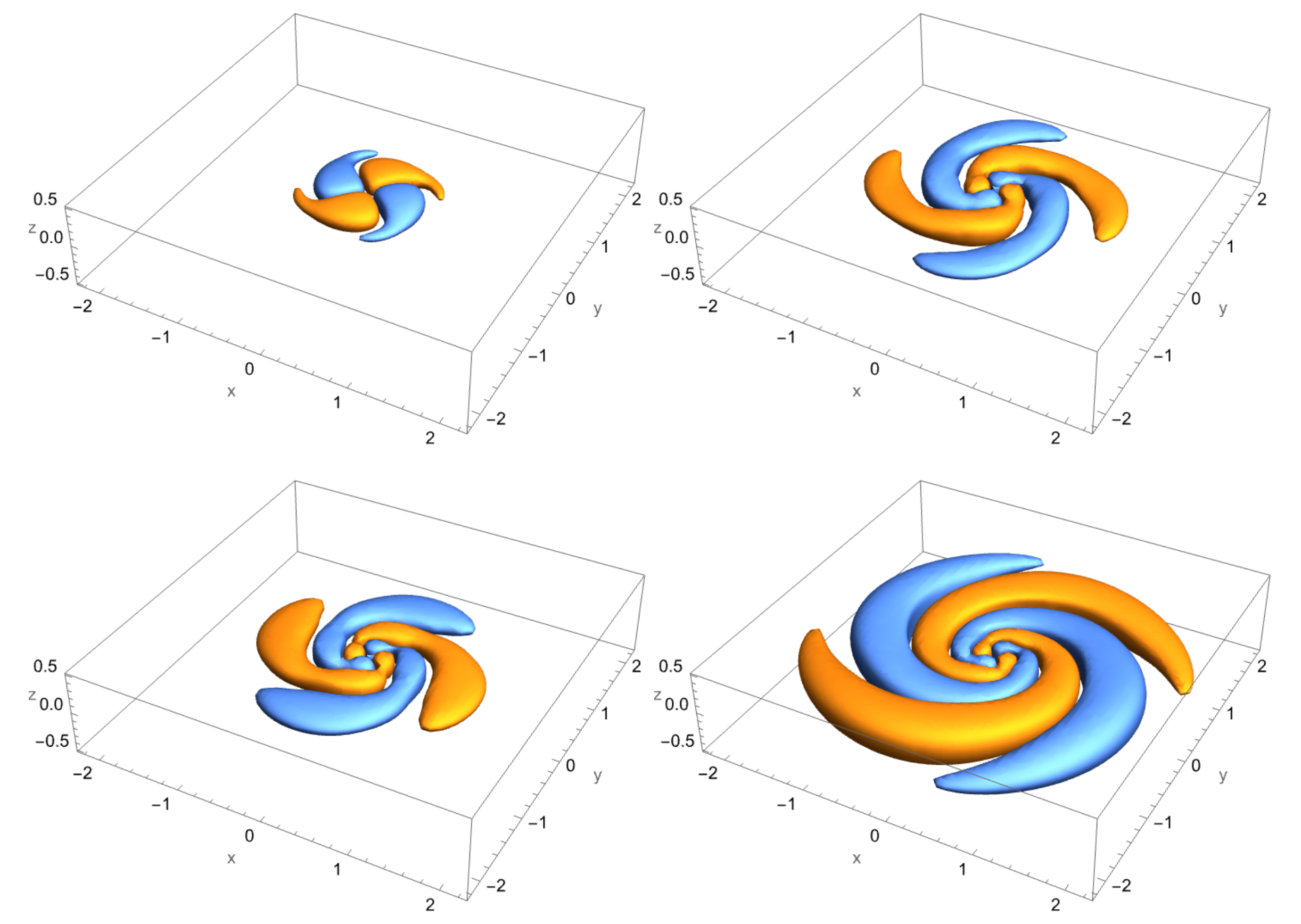}
   \caption{Shapes of the maximally rotating modes obtained in Fig.~\ref{fig:modes_subdominant}. \textit{Top left panel:} bar  mode 1. \textit{Top right panel:} bar  mode 2. \textit{Bottom left panel:} bar  mode 3. \textit{Bottom right panel:} bar  mode 4. The modes exhibit a spiral structure characteristic of the rapidly growing bar modes in the flattened Kuzmin--Kutuzov cluster, with each successive subdominant modes appearing increasingly wound and larger scale.}
   \label{fig:bar_modes_compare}
\end{figure*}

Let us extend the growth rates of the bending and bar-growing modes into their subdominant branches in Fig.~\ref{fig:modes_subdominant}. 
The bending mode approaches zero for the maximally rotating cluster, while the bar-growing modes exhibit the opposite behavior. This establishes a threshold at which highly flattened clusters are the least prone to instability.  For highly flattened clusters, we note the presence of multiple bar modes -- whose shapes are given in Fig.~\ref{fig:bar_modes_compare} -- whose hierarchy appears to depend on the rotation of the cluster.

\end{document}